\documentclass[iop]{emulateapj}

\usepackage{appendix,natbib}
\usepackage{subfigure}
\usepackage[backref,colorlinks=true,citecolor=blue,linkcolor=blue,urlcolor=blue]{hyperref}
\usepackage[all]{hypcap}

\usepackage{amsmath}
\usepackage{appendix}

\newcommand{\halpha}{H\ensuremath{\alpha}}
\newcommand{\hbeta}{H\ensuremath{\beta}}
\newcommand{\um}{$\mu$m}
\newcommand{\ratio}{$L_{\rm 7.7}/L_{\rm IR}$}
\newcommand{\lir}{$L_{\rm IR}$}
\def\o32{{O$_{\rm 32}$}}
\def\msun{{\rm M_\odot}}
\def\lsun{{\rm L_\odot}}
\def\sfr{{\rm SFR$_{\rm{\halpha,\hbeta}}$}}

\slugcomment{DRAFT: \today, Accepted to ApJ}

\begin{document}

\title{The {MOSDEF} Survey: Metallicity Dependence of PAH Emission at High Redshift \\and Implications for 24\,{\um}-inferred IR Luminosities and Star Formation Rates at $\lowercase{z}\sim 2$}

\author{\sc Irene Shivaei\altaffilmark{1,5}, Naveen A. Reddy\altaffilmark{1,6}, Alice E. Shapley\altaffilmark{2}, Brian Siana\altaffilmark{1}, Mariska Kriek\altaffilmark{3}, Bahram Mobasher\altaffilmark{1}, Alison L. Coil\altaffilmark{4}, William R. Freeman\altaffilmark{1}, Ryan L. Sanders\altaffilmark{2}, Sedona H. Price\altaffilmark{3}, Mojegan Azadi\altaffilmark{4}, Tom Zick\altaffilmark{3}}

\altaffiltext{1}{Department of Physics \& Astronomy, University of California, Riverside, CA 92521, USA}
\altaffiltext{2}{Department of Physics \& Astronomy, University of California, Los Angeles, CA 90095, USA}
\altaffiltext{3}{Astronomy Department, University of California, Berkeley, CA 94720, USA}
\altaffiltext{4}{Center for Astrophysics and Space Sciences, University of California, San Diego, La Jolla, CA 92093, USA}
\altaffiltext{5}{NSF Graduate Research Fellow}
\altaffiltext{6}{Alfred P. Sloan Research Fellow}

\begin{abstract}

We present results on the variation of 7.7\,{\um} Polycyclic Aromatic Hydrocarbon (PAH) emission in galaxies spanning a wide range in metallicity at $z\sim 2$. For this analysis, we use rest-frame optical spectra of 476 galaxies at $1.37\leq z\leq 2.61$ from the MOSFIRE Deep Evolution Field (MOSDEF) survey to infer metallicities and ionization states. {\em Spitzer}/MIPS 24{\um} and {\em Herschel}/PACS 100 and 160\,{\um} observations are used to derive rest-frame 7.7\,{\um} luminosities ($L_{7.7}$) and total IR luminosities ({\lir}), respectively. 
We find significant trends between the ratio of $L_{7.7}$ to {\lir} (and to dust-corrected SFR) and both metallicity and [O{\sc III}]/[O{\sc II}] ({\o32}) emission-line ratio. The latter is an empirical proxy for the ionization parameter. These trends indicate a paucity of PAH emission in low metallicity environments with harder and more intense radiation fields. 
Additionally, {\ratio} is significantly lower in the youngest quartile of our sample (ages of $\lesssim 500$\,Myr) compared to older galaxies, which may be a result of the delayed production of PAHs by AGB stars. The relative strength of $L_{7.7}$ to {\lir} is also lower by a factor of $\sim 2$ for galaxies with masses $M_*<10^{10}\msun$, compared to the more massive ones.
We demonstrate that commonly-used conversions of $L_{7.7}$ (or 24\,{\um} flux density; $f_{24}$) to {\lir} underestimate the IR luminosity by more than a factor of 2 at $M_*\sim 10^{9.6-10.0}\,\msun$. We adopt a mass-dependent conversion of $L_{7.7}$ to {\lir} with {\ratio}$=0.09$ and 0.22 for $M_*\leq 10^{10}$ and $> 10^{10}\,\msun$, respectively. Based on the new scaling, the SFR-$M_*$ relation has a shallower slope than previously derived. Our results also suggest a higher IR luminosity density at $z\sim 2$ than previously measured, corresponding to a $\sim 30\%$ increase in the SFR density.

\end{abstract}


\maketitle

\section{Introduction}

Emission from single-photon, stochastically-heated Polycyclic Aromatic Hydrocarbon (PAH) molecules dominates the mid-infrared (mid-IR) spectra of star-forming galaxies. The origin and properties of these molecules have been the subject of many studies in the literature, most of which have focused on local galaxies \citep[][and references therin]{tielens08}. 

It is crucial to fully understand how PAH emission depends on the physical conditions of the interstellar media (ISM) as the PAH emission contribution to total IR emission of galaxy may vary significantly from $\sim 1$ to 20\% in different environments \citep[][among many others]{smith07,dale09}.
There is evidence that PAHs are less abundant in metal-poor environments in the local universe \citep[e.g.,][]{normand95,calzetti07,draine07b,smith07,engelbracht05,hunt10,cook14}.
The physical explanation of this observation is a subject of much debate -- whether the decrease in the PAH emission at low metallicity is directly driven by metallicity or some other property of such environments is still unknown.
The most favored explanation is destruction of PAH molecules by hard UV radiation in low-metallicity environments due to reduced shielding by dust grains \citep[e.g.,][]{voit92,madden06,hunt10,sales10,khramtsova13,magdis13}. Other possibilities have also been discussed in the literature: that small PAH carriers are destroyed in low-metallicity environments through sputtering \citep[][and references therein]{hunt11}, that PAH formation and destruction mechanisms depend on dust masses, which in turn correlate with metallicity \citep{seok14}; and that the PAH-metallicity trend is a consequence of the PAH-age correlation \citep{galliano08}. The latter scenario is offered based on the assumption that the contribution of AGB stars -- as the purported primary origin of PAH molecules -- to ISM chemical enrichment increases with age.

The PAH emission features span from 3\,{\um} to 17\,{\um}, with the one at 7.7{\um} being the strongest \citep[contributing $\sim$ 40--50\% of the total PAH luminosity;][]{tielens08,hunt10}. At $z\sim2$, the 24\,{\um} filter of the {\em Spitzer}/MIPS instrument traces this feature. Due to the high sensitivity of MIPS, many high-redshift studies adopt the 24\,{\um} flux as an indicator of total IR luminosity ($L(8-1000\mu{\rm m})\equiv L_{\rm IR}$) and star-formation rate \citep[SFRs; e.g.,][]{ce01,reddy06a,daddi07a,wuyts08,reddy10,shivaei15a}. However, the metallicity and ionization state dependence of the PAH-to-{\lir} ratio of distant galaxies have not been studied in detail. In high-redshift studies, a single conversion from 24\,{\um} flux (or rest-8\,{\um} luminosity) to {\lir} is typically assumed for galaxies with a range of different metallicities and stellar masses \citep[e.g.,][]{wuyts08,wuyts11a,elbaz11,reddy12b,whitaker14b}.
Possible variations of the relative strength of PAH emission to {\lir} with metallicity (and as a consequence with stellar mass) can potentially alter the results of studies that rely on 24\,{\um} flux to infer {\lir} or SFR -- for example, those that investigate dust attenuation parameterized by {\lir}$/L_{{\rm UV}}$ \citep[e.g.,][]{reddy12b,whitaker14b}, or those that utilize bolometric SFRs (i.e., SFR$_{\rm IR}+$SFR$_{\rm UV}$) to explore relations such as the SFR-$M_*$ relation \citep[e.g.,][]{daddi07a,wuyts11b,fumagalli14,tomczak16}.

With the large and representative dataset of the MOSFIRE Deep Evolution Field (MOSDEF) survey \citep{kriek15}, we are in a unique position to investigate, for the first time, the dependence of PAH intensity (defined as the ratio of 7.7\,{\um} luminosity to SFR or {\lir}) on the ISM properties of high-redshift galaxies. The MOSDEF survey provides us with near-IR spectra of galaxies at $1.37\leq z\leq 2.61$, from which we calculate spectroscopic redshifts and estimate gas-phase metallicities and ionization states. We use mid- and far-IR photometric data from {\em Spitzer}/MIPS 24\,{\um} and {\em Herschel}/PACS 100 and 160\,{\um} to measure PAH emission and total IR luminosities, respectively.

Our study includes galaxies over a broad range of stellar masses ($M_*\sim 10^9-10^{11.5}\,\msun$), SFRs ($\sim 1-200\,\msun\,{\rm yr^{-1}}$), and metallicities ($\sim 0.2-1\,Z_{\odot}$). Ultimately, we quantify how the conversions between rest-frame 7.7\,{\um} and both SFR and {\lir} depend on metallicity and stellar mass. These scaling relations are important for deriving unbiased estimates of total IR luminosities and obscured SFRs based on observations of the PAH emission in distant galaxies -- such observations will be possible for larger numbers of high-redshift galaxies with {\em JWST}/MIRI \citep{shipley16}.

The outline of this paper is as follows. In Section~\ref{sec:data}, we introduce the MOSDEF survey and describe our measurements including line fluxes, stellar masses, SFRs,  IR photometry, and the IR stacking method. In Section~\ref{sec:pah_ism}, we constrain the dependence of $L_{7.7}/{\rm SFR}$ and {\ratio} on metallicity and ionization state. The PAH intensity as a function of age is explored in Section~\ref{sec:pah_age}. Implications of our results for the studies of the SFR-$M_*$ relation and the IR luminosity density at $z\sim 2$ are discussed in Section~\ref{sec:implications}. In Section~\ref{sec:discussion}, we briefly discuss the possible physical mechanisms driving the PAH-metallicity correlation. Finally, the results are summarized in Section~\ref{sec:conclusion}. Throughout this paper, line wavelengths are in vacuum and we assume a \citet{chabrier03} initial mass function (IMF). A cosmology with $H_0=70\,{\rm km\,s^{-1}\,Mpc^{-1}}, \Omega_{\Lambda}=0.7, \Omega_{{\rm m}}=0.3$ is adopted.

\section{Data}
\label{sec:data}

\subsection{The MOSDEF Survey}
\label{sec:data_mosdef}
During the MOSDEF survey, we obtained rest-frame optical spectra of $\sim 1500$ galaxies with the MOSFIRE spectrograph on the Keck~I telescope \citep{mclean12}. The parent sample was selected based on {\em H}-band magnitude in three redshift ranges: $z=1.37-1.70$, $2.09-2.61$, and $2.95-3.80$, down to $H=24.0$, 24.5, and 25.0 mag, respectively. These redshift ranges were adopted to ensure coverage of the strong optical emission lines ([O{\sc II}], [O{\sc III}], {\hbeta}, {\halpha}) in the {\em JHK} bands. The MOSDEF survey was conducted in the five CANDELS fields: AEGIS, COSMOS, GOODS-N, GOODS-S, and UDS \citep{grogin11,koekemoer11}. These fields are also covered by the HST/WFC3 grism observations of the 3D-HST survey \citep{skelton14,momcheva16}. The details of the MOSDEF survey, observing strategy, and data reduction are reported in \citet{kriek15}.

\begin{figure}[tbp]
	\centering
	\subfigure{
		\includegraphics[width=.45\textwidth]{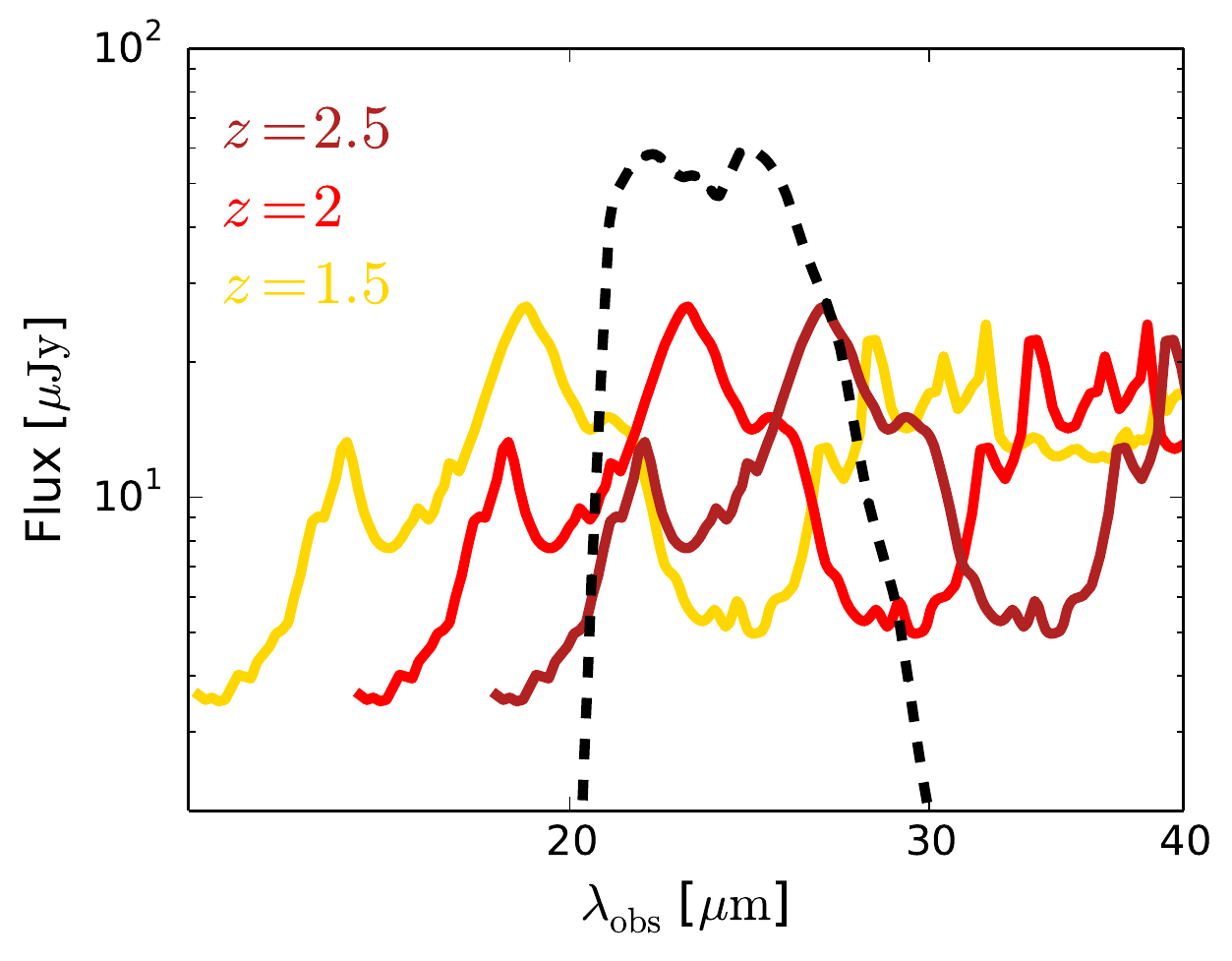}}
		\caption{The {\em Spitzer}/MIPS 24{\um} filter (black dashed line) superimposed on the observed mid-IR spectrum of a galaxy at $z=1.5$ (yellow), $z=2$ (light red), and $z=2.5$ (dark red). The mid-IR spectrum is adopted from \citet{rieke09} for a galaxy with {\lir} $=10^{11}\lsun$.
		}		
	\label{fig:filter}
\end{figure}

We limit our study to the first two redshift bins ($1.37\leq z\leq 1.70$ and $2.09\leq z\leq 2.61$), where the MIPS 24\,{\um} filter covers the rest-frame 7.7\,{\um} PAH feature (Figure~\ref{fig:filter}). Objects with active galactic nucleus (AGN) contamination are removed from our study, based on their X-ray emission, IRAC colors (using the \citealt{donely12} criteria), and/or [N{\sc II}]/{\halpha} line ratios ([N{\sc II}]/{\halpha} $> 0.5$; \citealt{coil15,azadi16}).

\subsection{Emission Line Measurements}
\label{sec:lines}

Emission line fluxes were estimated by fitting Gaussian functions to the line profiles. Uncertainties in the line fluxes were derived by perturbing the spectrum of each object according to its error spectrum and measuring the line fluxes in these perturbed spectra \citep[see][]{kriek15,reddy15}. The {\halpha} line and [N{\sc II}] doublet were fit with three Gaussian functions and the [O{\sc II}] doublet was fit with two Gaussians. To account for the loss of flux outside of the spectroscopic slits, corrections were applied by normalizing the spectrum of a slit star in each observing mask to match the 3D-HST total photometric flux \citep{skelton14}. Additionally, we used the {\em HST} images of our resolved galaxy targets to estimate and correct for differential flux loss relative to that of the slit star \citep{kriek15}. {\halpha} and {\hbeta} fluxes were further corrected for underlying Balmer absorption as determined from the best-fit stellar population models to the broadband photometry (Section~\ref{sec:data_sfr}).

We use the [N{\sc II}]$\lambda 6585$/{\halpha} (N2) and ([O{\sc III}]$\lambda 5008$/{\hbeta})/([N{\sc II}]$\lambda 6585$/{\halpha}) (O3N2) indicators to derive gas-phase oxygen abundances (metallicities) based on the empirical calibrations of \citet{pp04} \citep[see][]{sanders15,shapley15}. As both N2 and O3N2 indicators include ratios of emission lines that are close in wavelength space, no dust correction is applied. Although there are concerns regarding biases in the metallicity indicators that use nitrogen \citep{shapley15,sanders16a}, the N2 and O3N2 indicators still distinguish the lower- and higher-metallicity galaxies, which is sufficient for the purpose of this study.
Furthermore, we use the [O{\sc III}]$\lambda\lambda 4960, 5008$/[O{\sc II}]$\lambda\lambda 3727, 3730$ ratio ({\o32}) as a proxy for ionization parameter. As [O{\sc III}]$\lambda 5008$ has a higher signal-to-noise (S/N) than [O{\sc III}]$\lambda 4960$, we assume a fixed [O{\sc III}]$\lambda 5008$/[O{\sc III}]$\lambda 4960$ line ratio of 2.98 \citep{storey00} to calculate the sum of [O{\sc III}]$\lambda 5008$ and [O{\sc III}]$\lambda 4960$. We correct [O{\sc III}] and [O{\sc II}] lines for dust extinction by using the Balmer decrement ({\halpha}/{\hbeta}) and assuming the \citet{cardelli89} extinction curve \citep{reddy15,shivaei15b}. Where necessary, we calculate {\o32} gas-phase metallicity based on the calibrations of \citealt{jones15}.

\subsection{Stellar Masses, Ages, and SFRs}
\label{sec:data_sfr}

Stellar masses and ages are derived by fitting rest-UV to near-IR photometry from 3D-HST \citep{skelton14,momcheva16} with \citet{bruzual03} models through a minimum $\chi^2$ method. The photometry is corrected for emission line contamination according to the MOSDEF spectra. For the SED fitting we assume a solar metallicity, a \citet{chabrier03} IMF, and an exponentially rising star-formation history \citep{reddy15}.
The latter is assumed because it has been shown that rising star formation histories best reproduce the observed SFRs at $z\sim 2$ \citep[e.g.,][]{wuyts11a,reddy12b}.
As described in \citet{reddy12b}, ages that are inferred from exponentially rising star formation histories are ambiguous and often larger than those derived from constant star formation histories. However, a majority (86\%) of our galaxies have very large characteristic timescales ($\tau=5000$\,Myr), which makes their star formation histories very similar to a constant one, along with well-defined ages.

We convert {\halpha} luminosities to SFRs by adopting the \citet{kennicutt98} relation, modified for the \citealt{chabrier03} IMF. The line luminosities are corrected for dust attenuation using the Balmer decrement and assuming the \citet{cardelli89} curve \citep{shivaei15b}. Hereafter, we refer to the dust-corrected {\halpha} SFR as {\sfr}.

\begin{figure}[tbp]
	\centering
	\subfigure{
		\includegraphics[width=.45\textwidth]{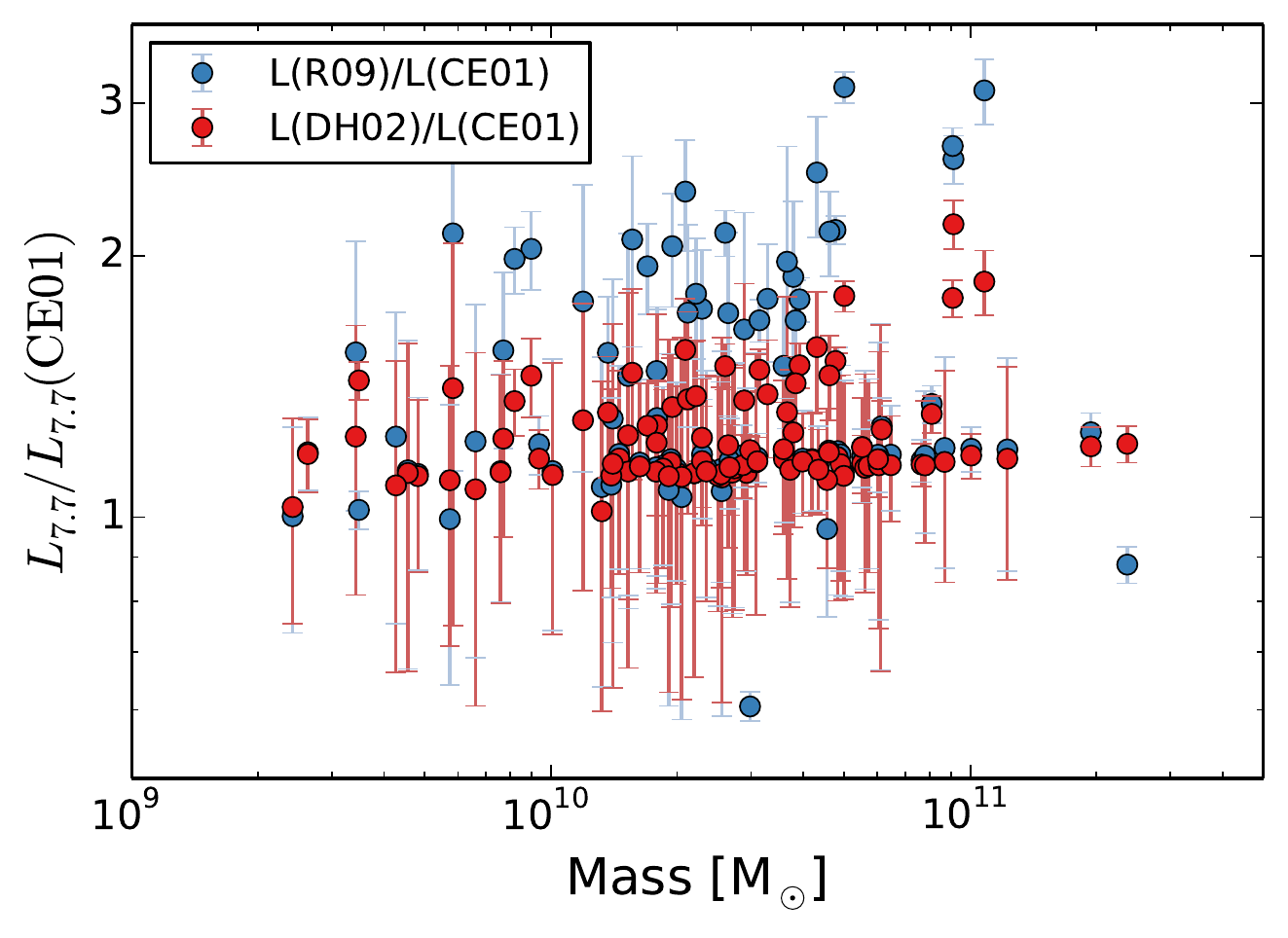}}
		\caption{Ratio of the rest-frame 7.7\,{\um} luminosities inferred from the IR templates of \citet[][red]{dh02} and \citet[][blue]{rieke09} to those inferred from the \citet{ce01} templates for 299 galaxies with robust redshifts.
		The 7.7\,{\um} luminosities are derived by fitting 24\,{\um} flux densities to the IR templates.
		For a majority of galaxies, the systematic bias caused by using different templates is not significant compared to the measurement uncertainties.
		}
	\label{fig:l8_models}
\end{figure}

\subsection{Mid- and Far-IR fluxes}
\label{sec:data_psf}
We perform scaled point-spread function (PSF) photometry on the {\em Spitzer}/MIPS and {\em Herschel}/PACS images in COSMOS, GOODS-N, GOODS-S, and AEGIS fields from multiple observing programs \citep[PI: M.~Dickinson;][]{dickinson07,elbaz11,magnelli13}. The accuracy of the measured fluxes is determined by adding simulated sources to the science images and recovering their fluxes in the same way as for real sources. Details of the photometry and simulations are described in \citet{shivaei16a}. In brief, we use the IRAC and MIPS priors down to a $3\sigma$ limit for MIPS and PACS images, respectively. A 40 by 40 pixel subimage is made around each target. In each subimage, the PSFs are scaled to match the prior objects and the target simultaneously. In this process, a covariance matrix is defined to determine the robustness of the fit, which is affected by confusion with nearby sources as described in \citet{reddy10}. Based on the total covariance value of the MIPS 24\,{\um} fits, we remove objects that have nearby companions from our analysis. To determine the noise and background level in each subimage, we fit PSFs to 20 random positions, ensuring that these positions are more than 1 FWHM away from any real sources. The average and standard deviation of the background fluxes are adopted as the background level and noise in the image, respectively. The typical measurement uncertainty for 24\,{\um}-detected objects is $\sigma_{24} \sim 15\%$. The majority ($\sim 80\%$) of these 24\,{\um}-detected objects are not detected in 100 and 160\,{\um} images.
Out of 406 non-AGN objects without any nearby sources (see above), 128 of them (32\%) are detected at 24\,{\um} with $S/N>3$, and only 13 (3\%) are detected with $S/N>3$ at both 24 and 100\,{\um}.

\subsection{IR Stacks}
\label{sec:data_irstack}

Due to the relatively shallow depth of the available far-IR data, we stack the PACS images to obtain a sufficiently high $S/N$, so that we can measure average IR luminosities. We also stack the {\em Spitzer}/MIPS 24\,{\um} images because of the high fraction of 24\,{\um}-undetected objects ($S/N<3$). 
In this analysis, all the stacks include both the detected and undetected objects.
The stacking technique is described below.

We extract 40$\times$40 pixel subimages centered on each target, and subtract all prior sources brighter than the detection limit of $S/N=3$, using the scaled PSF method described in Section~\ref{sec:data_psf}. We then combine these residual images by making an inverse-SFR-weighted average stack as follows. We normalize each 24\,{\um} image by its {\sfr} and divide the sum of the normalized images by the sum of the weights (i.e., 1/SFR).
We adopt inverse-SFR-weighted average stacks because ultimately we are interested in investigating the average of $f_{24}$/SFR and $L_{7.7}$/SFR ratios (Section~\ref{sec:pah_sfr_ism}). As we do not have direct far-IR detections for a majority of our galaxies, we can not construct inverse-{\lir}-weighted average stacks. As a result, we use a 3$\sigma$-clipped mean or a median stack of 24, 100, and 160\,{\um} images for the $f_{24}$/{\lir} and $L_{7.7}$/{\lir} analyses (Section~\ref{sec:pah_ir_ism}). A comparison of these different stacking methods is presented in Appendix~\ref{sec:appA}.

We calculate the stacked flux by performing aperture photometry on the stacked image.
Based on the measured $S/N$ of various aperture sizes, we adopt a 4-pixel radius aperture. The amount of light falling outside of the 4-pixel aperture is calculated using the PSF, and the fluxes are corrected accordingly. Similar to the PSF fitting technique, we measure background fluxes in 20 random positions to determine the noise (i.e., the uncertainty in the stacked 24\,{\um} flux). There is little variation in the aperture corrections from field-to-field ($\lesssim 5\%$), which is negligible compared to the measurement errors.

\subsection{$\nu L_{\nu}$(7.7\,{\um}) and {\lir}}
\label{sec:data_ir}

The {\em k}-correction required for converting 24\,{\um} flux densities to rest-frame 7.7\,{\um} luminosities is computed using the \citet[][hereafter, CE01]{ce01} templates. First, we shift the templates to the redshift of each object (or median/weighted-average redshift of the stack), and then convolve the models with the 24\,{\um} filter transmission curve. The best-fit model is determined through a least-$\chi^2$ method and $\nu L_{\nu}(7.7\mu{\rm m})$ ($L_{7.7}$, in units of $\lsun$) is extracted from the best-fit model. 

To quantify the systematic biases from using different IR templates, we measure $L_{7.7}$ from the best-fit \citet{dh02} and \citet{rieke09} models and compare them with those of CE01 (Figure~\ref{fig:l8_models}). Adopting the \citet{dh02} and \citet{rieke09} models would systematically increase the calculated $L_{7.7}$ by $\sim 0.06$ and $\sim 0.10$\,dex, respectively. 
For a small fraction (4\%) of galaxies, the \citet{rieke09} $L_{7.7}$ is larger than the CE01 $L_{7.7}$ by $>0.3$\,dex. However, these galaxies do not have systematically different masses or metallicities compared to that of the rest of the sample. Therefore, their $L_{7.7}$ offsets do not affect the mass and metallicity trends in this study.

Total IR luminosities ({\lir}, integrated from 8 to 1000\,{\um}) are measured from the best-fit CE01 models to PACS 100 and 160{\um} stacks. We measure {\lir} errors by perturbing the 100 and 160{\um} flux densities of each stack by their measurement uncertainties 10,000 times and calculating the 68\% confidence intervals from these realizations. Furthermore, we verify that the best-fit models of \citet{dh02} alter the inferred IR luminosities by only $\sim 5\%$, which is negligible compared to the typical {\lir} measurement uncertainties of $\sim 10-20\%$.

\begin{figure}[tbp]
	\subfigure{
	\centering
		\includegraphics[width=.9\columnwidth]{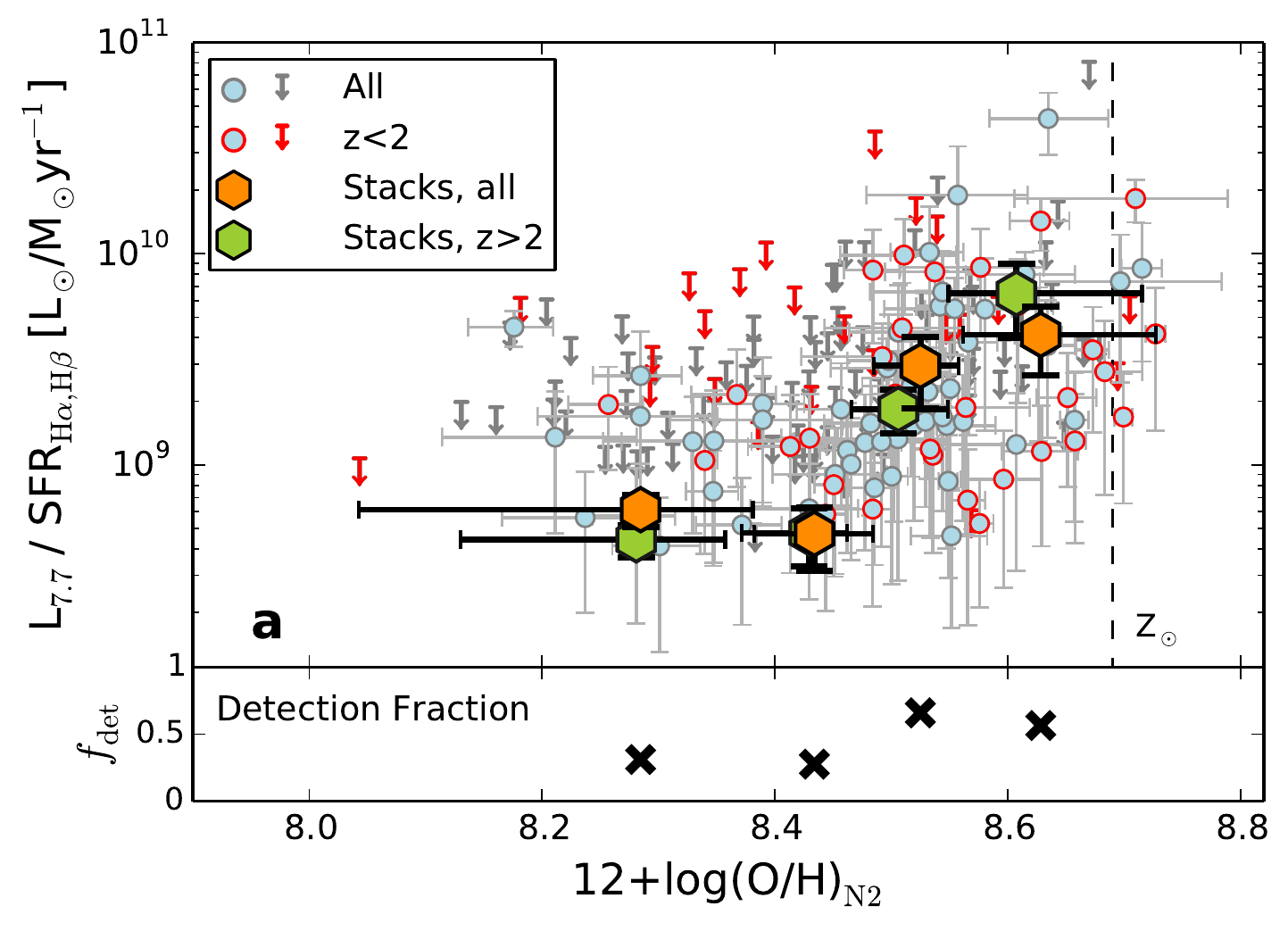}}
	\\
	\subfigure{
	\centering
		\includegraphics[width=.9\columnwidth]{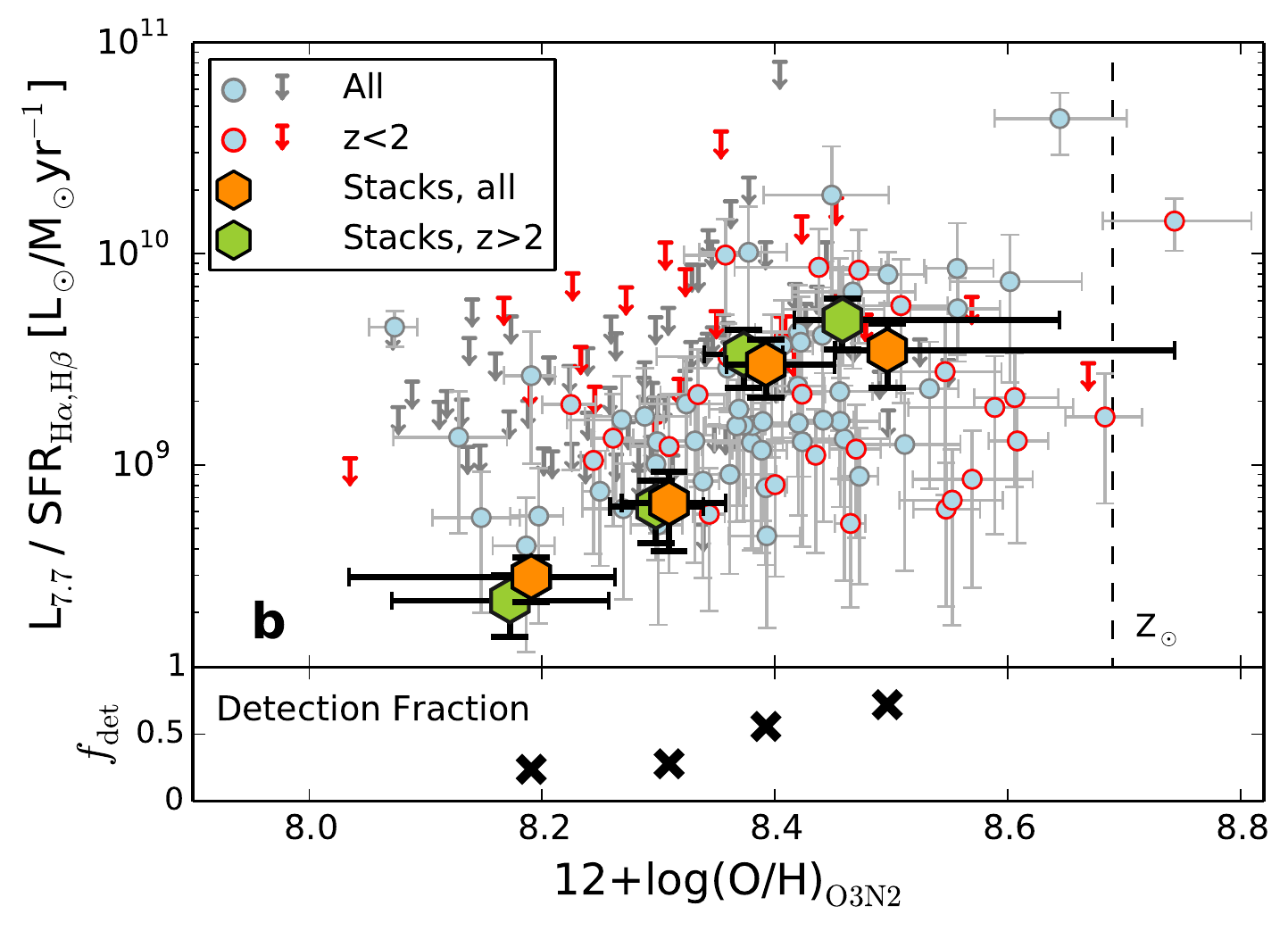}}
	\\
	\subfigure{
	\centering
		\includegraphics[width=.9\columnwidth]{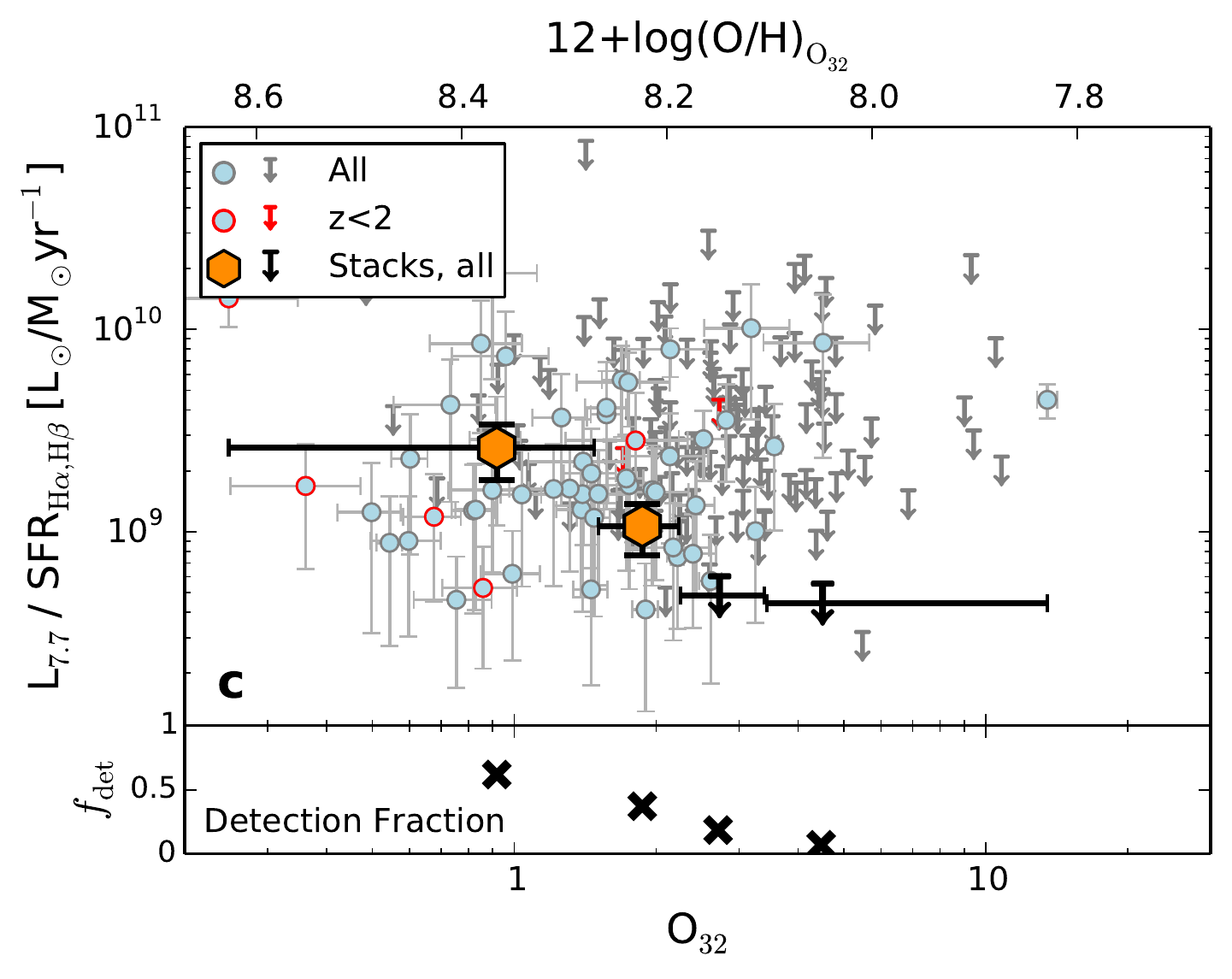}}
		\caption{Ratio of 7.7\,{\um} luminosity to dust-corrected {\sfr} as a function of (a) N2 metallicity, (b) O3N2 metallicity, and (c) {\o32} ratio. The {\o32} gas-phase metallicity (based on calibrations of \citealt{jones15}) is also displayed in panel (c).
		The plots show galaxies that have detections ({\it S/N}$>3$) in all of the diagnostic emission lines used in each plot: panel (a) has objects detected in {\halpha}, {\hbeta}, and [N{\sc II}] (total of 187 objects), panel (b) has objects detected in {\halpha}, {\hbeta}, [N{\sc II}], and [O{\sc III}] (172 objects), and objects in panel (c) are detected in {\halpha}, {\hbeta}, [O{\sc III}], and [O{\sc II}] (171 objects).
		AGNs and objects with 24\,{\um} nearby neighbors are removed.
		Small circles and arrows indicate individual detections and 3\,$\sigma$ upper limits for non-detections at 24\,{\um}, respectively. The 24\,{\um} detection fraction in each bin is shown in the bottom panels.
		Red symbols show galaxies at $1.37\leq z\leq 2.0$.
		The $L_{7.7}$/{\sfr} stacks are performed in bins of the quantity on the horizontal axis and include all the 24\,{\um} detected and undetected galaxies in each bin. The horizontal error bars show the width of the bins. 
		In panel (c), there are only 8 objects at $z<2$, because [O{\sc II}] is not covered by the MOSFIRE Y filter at $z\leq 1.6$. 
		In panels (a) and (b), solar metallicity ($12+\log({\rm O/H})=8.69$, \citealt{asplund09}) is indicated with a dashed line.
		Stack values are listed in Table~\ref{tab:pah_ism}.
		}
	\label{fig:pah_ism}
\end{figure}

\section{PAH and ISM properties}
\label{sec:pah_ism}

In this section, we explore variations of the 7.7\,{\um} PAH intensity in different ISM conditions, specifically with different gas-phase metallicities and ionization parameters. As a consequence of the correlation between metallicity and dust, we expect {\lir} and SFR to correlate with metallicity such that metal-rich galaxies have higher SFRs and {\lir}. We quantify the PAH emission intensity by normalizing $L_{7.7}$ to either {\sfr} (Section~\ref{sec:pah_sfr_ism}) or {\lir} (Section~\ref{sec:pah_ir_ism}). 

The fractional contribution of the stellar and dust continuum from very small grains (VSGs) to the 7.7\,{\um} PAH emission varies in star-forming galaxies. 
Using a sample of local starburst galaxies, \citet{engelbracht08} showed that the stellar fractional contribution in the IRAC 8\,{\um} band is 30 to 4\% for $12+\log({\rm O/H})\sim 7.4$ to 8.7, respectively. Galaxies in our sample have $12+\log({\rm O/H})>8$, suggesting a stellar fraction of $< 10\%$. Moreover, following the redshift evolution of SFR-$M_*$ relation from $z\sim 0$ to 2 \citep[e.g.,][]{whitaker14b}, our galaxies have typically higher specific SFRs, and hence we expect lower stellar contamination in their mid-IR bands relative to local galaxies of the same stellar mass.
Additionally, spectra of lensed high-redshift galaxies indicate a weak presence of mid-IR continuum \citep{rigby08,siana09,fadely10} and high PAH emission contribution to the broadband 24\,{\um} filter at $z\sim 2$ \citep{smith07}. Therefore, we conclude that the continuum contribution to our $L_{7.7}$ measurements is negligible. Future observations with {\em JWST}/MIRI will help to clarify the stellar and dust continuum contribution to the aromatic bands in the spectra of high-redshift galaxies. Compared to star-forming galaxies, AGNs typically have fainter PAH emission and a higher (non-negligible) fractional contribution of thermal and stellar continuum to the mid-IR emission \citep{smith07}. We removed the known AGNs from our sample (Section~\ref{sec:data_mosdef}). 

\subsection{$L_{7.7}$/SFR and ISM properties}
\label{sec:pah_sfr_ism}

Due to the relatively shallow depth and poor spatial resolution of the far-IR data, obtaining robust individual far-IR measurements for a representative number of galaxies in our sample is not possible. Instead, we rely on our individual measurements of {\sfr} (Section~\ref{sec:data_sfr}). It was shown in \citet{shivaei16a} that {\sfr} accurately traces SFRs up to $\sim 300\,\msun\,{\rm yr}^{-1}$, when compared with those from the IR.

We calculate inverse-{\sfr}-weighted average stacks of 24\,{\um} images (see Section~\ref{sec:data_irstack}) in four bins of metallicity with a roughly equal number of galaxies in each bin. The stacked 24\,{\um} flux density is converted to $L_{7.7}$ using the CE01 IR templates (Section~\ref{sec:data_ir}), and then divided by the weighted-average {\sfr}. Dividing the inverse-{\sfr}-weighted average $L_{7.7}$ to weighted average {\sfr} is mathematically equivalent to calculating an average of the $L_{7.7}$/{\sfr} ratios in each bin (Appendix~\ref{sec:appA}).
Properties of the stacks are listed in Table~\ref{tab:pah_ism}. Figure~\ref{fig:pah_ism}(a,b) shows $L_{7.7}$ to {\sfr} ratio as a function of N2 and O3N2 metallicities 
\footnote{Based on a small sample of three galaxies at $z\sim 2$, one lensed galaxy at $z=1.4$, and nine $z\simeq 0.2$ green pea galaxies, \citet{steidel14} showed that the O3N2 index indicates a slightly better agreement with the direct $T_e$ measurements of $12+\log(O/H)$, relative to those computed using the N2 index.} (Section~\ref{sec:lines}).
The uncertainties in $L_{7.7}$/SFR ratios are calculated by adding the $L_{7.7}$ and SFR fractional uncertainties in quadrature. There is a clear increase of $L_{7.7}$/SFR with increasing metallicity. 
Considering the small fraction of 24\,{\um}-detected objects in the two lowest metallicity bins, it is expected that the average $L_{7.7}$/{\sfr} ratios, which include all the detected and undetected 24\,{\um} objects in the corresponding bins, will lie below the few individually 24\,{\um}-detected sources.
We consider separately the trends for the full redshift range, and for $z>2$ alone. The $L_{7.7}$/SFR ratio increases by a factor of $\sim 10$ ($\sim 15$) from the lowest to the highest metallicity bin in stacks of galaxies at $1.37\leq z\leq 2.61$ ($2.0\leq z\leq 2.61$).

We investigate the potential incompleteness of our samples due to the requirement of detecting the [N{\sc II}] and [O{\sc III}] lines for N2 and O3N2 metallicties.
The fractions of galaxies detected in [N{\sc II}] ([O{\sc III}]) for the four bins of stellar mass indicated in Table 2 are 0.28, 0.56, 0.72, and 0.93 (0.28, 0.55, 0.68, and 0.83) proceeding from low to high stellar masses. As expected, more galaxies with undetected emission lines are missed from the N2 and O3N2 samples as we go to lower masses. However, in each mass bin, the distributions of the 24\,{\um} {\it S/N} and $f_{24}$ of the detected-line subsamples are very similar to that of the full sample (including detected and undetected [N{\sc II}] and [O{\sc III}] lines).
Additionally, in Section~\ref{sec:sfr-mass} we show the $L_{7.7}$ and {\lir} stacks in bins of stellar mass regardless of the emission line detection. Both the $L_{7.7}$/SFR and {\ratio} ratios are suppressed at low masses, which correspond to low metallicities according to the mass-metallicity relation \citep{tremonti04}. We conclude that the sample incompleteness at low N2 and O3N2 metallicities is unlikely to be the dominant factor in driving the $L_{7.7}$/SFR and {\ratio} trends with metallicity.

Another important trend is the anti-correlation between the PAH intensity and {\o32} ratio. {\o32} is used as an empirical proxy for ionization parameter, which is in turn anti-correlated with metallicity (e.g., \citealt{perez-montero14}, see also Figure 12 in \citealt{sanders16a} for the relation of {\o32} with O3N2 and N2 in MOSDEF). The weighted-average stacks of 24\,{\um} images in the two highest {\o32} bins yield non-detections.
In addition, objects with detected and undetected 24\,{\um} emission are distinctly separated towards lower and higher {\o32}, respectively. 
The fraction of 24\,{\um}-undetected objects increases significantly from 38\% in the lowest {\o32} bin (median {\o32} $=0.9$) to 91\% in the highest bin (median {\o32} $=4.5$), while the 24\,{\um} non-detection fraction changes from $\sim 30\%$ at $12+\log({\rm O/H})\sim 8.6$ (8.5) to $\sim 60-70\%$ at $12+\log({\rm O/H})\sim 8.3$ (8.2) for N2 (O3N2) metallicities (refer to the bottom panels in Figure~\ref{fig:pah_ism}).
These trends suggest that the PAH intensity is strongly affected by the ionization parameter. We will return to this point in Section~\ref{sec:discussion}.

Excluding the 24\,{\um}-undetected objects from the analysis leads to a significantly biased sample. However, to explore the variations of 24\,{\um}-bright sources with {\o32}, we stack only the 24\,{\um}-detected objects in the same four bins of {\o32} as before. The results indicate that the $L_{7.7}$/SFR ratios of 24\,{\um}-bright sources are almost independent of {\o32} within the uncertainties ($L_{7.7}$/SFR $\sim 2.5\times10^{9}\,\lsun/\msun {\rm yr}^{-1}$ in all {\o32} bins.)

The PAH-metallicity correlation found at $z\sim 2$ is consistent with the trend observed at $z\sim 0$ \citep[][among others]{engelbracht05,draine07b,galliano08,hunt10}. 
In local galaxies, there is a paucity of PAH emission at $12+\log({\rm O/H})\lesssim 8.1-8.2$ \citep{engelbracht05,wu06,draine07b}. In Figure~\ref{fig:pah_ism}(c) we see the same threshold in the {\o32} plot at $z\sim 2$: there is a sharp difference in the $L_{7.7}$/{\sfr} ratio below and above ${\rm O}_{32}\sim 2$, which corresponds to $12+\log({\rm O/H)_{O_{32}}}\sim 8.2$ based on the metallicity calibrations of \citet{jones15}. The threshold is at $12+\log({\rm O/H)_{N2}}\sim 8.5$ for the N2 metallicity calibration.
We will discuss this observed threshold in more detail in Section~\ref{sec:discussion}.

Additionally, we examine possible variations of $L_{7.7}$ to {\sfr} (and to {\lir}) ratio in the BPT diagram ([O{\sc III}]/{\hbeta} vs. [N{\sc II}]/{\halpha}; \citealt{bpt81}), as the position of galaxies below and above the BPT locus is known to be sensitive to the hardness of ionizing radiation \citep{kewley13}. We use the locus of $z\sim 2$ star-forming galaxies in the BPT diagram from \citet{shapley15} and stack 24\,{\um} images of galaxies below and above the locus.
Surprisingly, we do not find any evidence for variation of $L_{7.7}$ intensity in this parameter space, as galaxies at $z\geq 2$ below and above the BPT locus have the same {\ratio} ratio of $0.20\pm 0.03$. {\em It is not clear why we do not see a significant change of the PAH intensity across the BPT diagram; it may be due to the small sample size.}

Our results directly show the correlation between the PAH emission and intensity of the radiation field as was speculated by \citet[][hereafter, E11]{elbaz11}. In \citet{elbaz11}, the reduced $L_8$/{\lir} ratio was attributed to star formation ``compactness'' and ``starburstiness'' characterized by the SFR surface density and specific SFR, respectively. These authors concluded that the weak PAH emission in compact starbursts compared to their typical star-forming counterparts is a consequence of an increased radiation field intensity in these galaxies, which is directly shown by our results of decreasing PAH intensity with {\o32}.
In other words, based on the assumption of a constant electron density (as electron density appears to be almost independent from other galaxy properties in MOSDEF, \citealt{sanders16a}), and the same ionizing spectrum, there is a tight correspondence between {\o32} and ionization parameter, such that increasing {\o32} reflects a more intense radiation field. This conclusion still holds even if we assume a different (harder) ionizing spectrum for the higher {\o32} bins, which would lower the inferred ionization parameter for a given {\o32} value \citep[see Figure 10 in][]{sanders16a}.

\begin{figure*}[tbp]
	\centering
	\subfigure{
		\includegraphics[width=.35\textwidth]{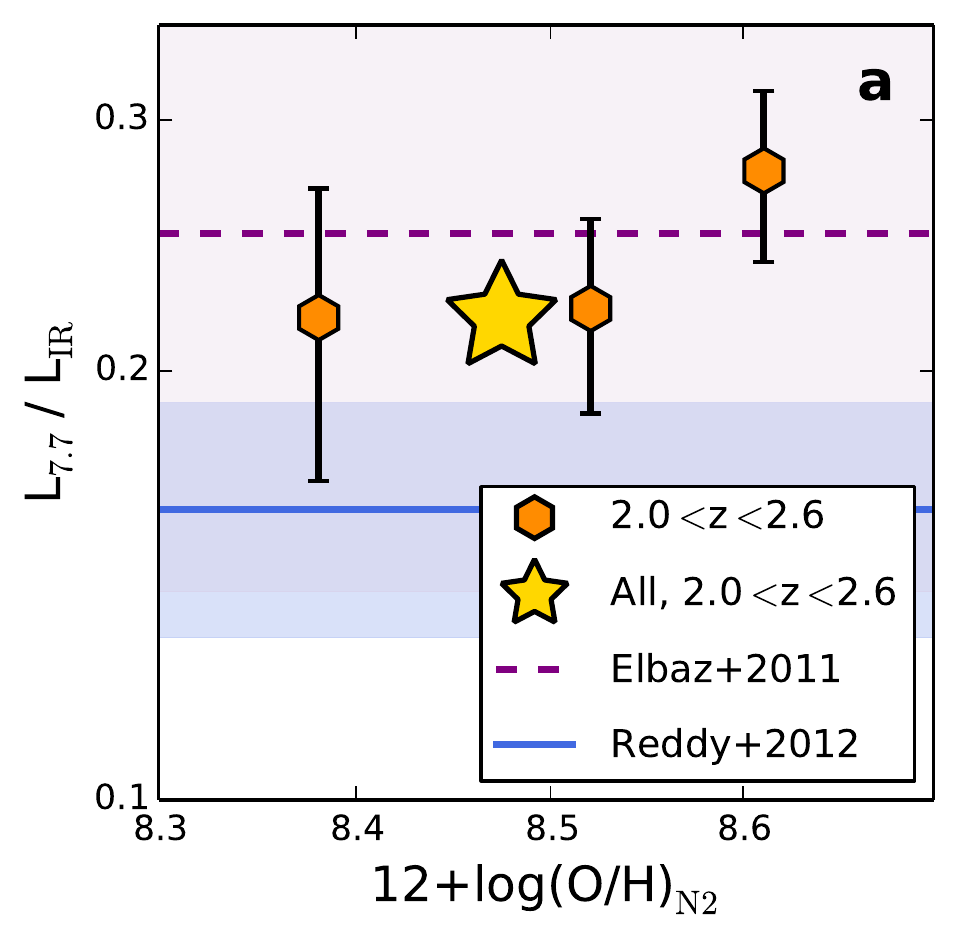}}
	\quad
	\centering
	\subfigure{
		\includegraphics[width=.35\textwidth]{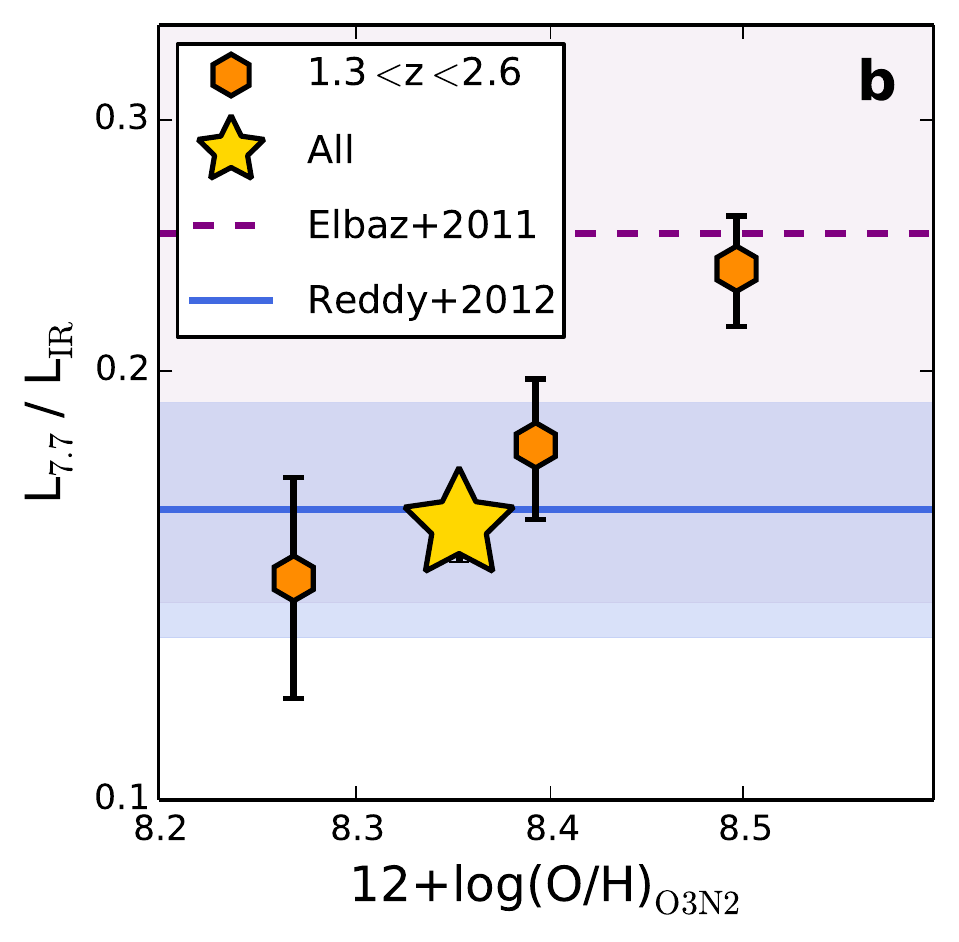}}
	\\
	\centering
	\subfigure{
		\includegraphics[width=.35\textwidth]{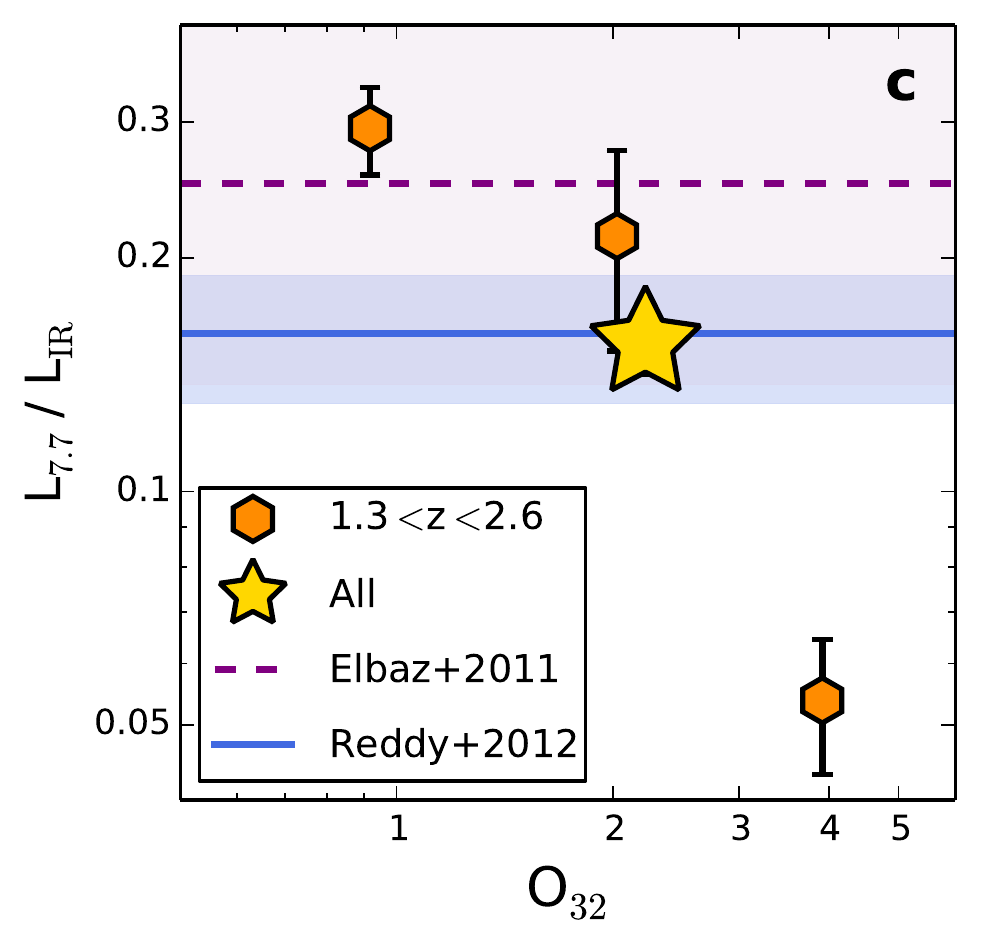}}
	\quad
	\centering
	\subfigure{
		\includegraphics[width=.36\textwidth]{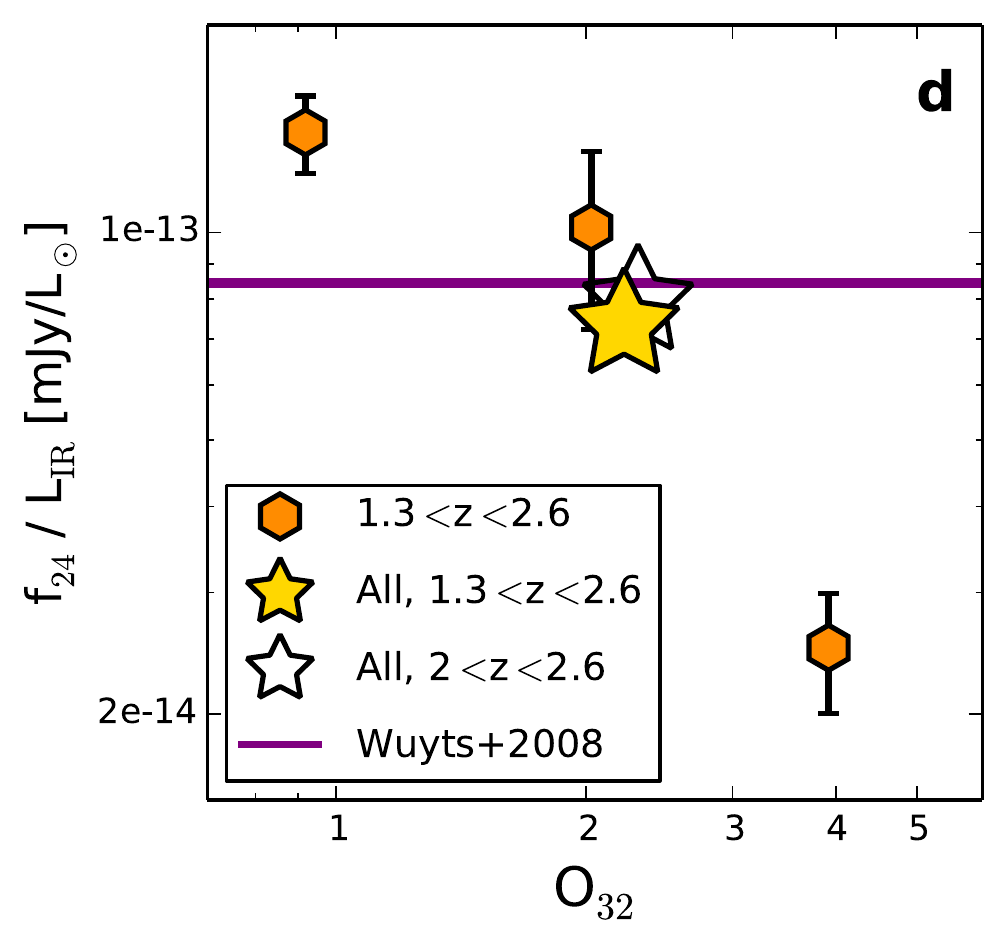}}
		\caption{Relative strength of 7.7\,{\um} luminosity to total IR luminosity as a function of N2 metallicity (a), O3N2 metallicity (b), {\o32} ratio (c), and ratio of 24\,{\um} flux density to total IR luminosity versus {\o32} (d).
		In order to gain sufficient {\it S/N} in PACS bands, the lowest metallicity bins and the highest {\o32} bin have twice the number of galaxies in other bins. 
		Yellow stars show stacks of all galaxies.
		For comparison, we show the {\ratio} conversions of E11 and R12 and the associated uncertainties with solid lines shaded regions, respectively. The $f_{24}/${\lir} ratio of W08 is plotted in panel d. The W08 ratio is redshift-dependent and the width of the purple line in plot (d) shows the range of values for the three bins.
		The N2 stacks are only performed for galaxies at $z>2$, because otherwise the last bin would be biased towards galaxies at $z<2.0$.
		Stacked values are listed in Table~\ref{tab:pah_ir}.
		}		
	\label{fig:pah_ir}
\end{figure*}

\subsection{{\ratio} and ISM properties}
\label{sec:pah_ir_ism}

Estimating the total IR luminosity independent of $f_{24}$ requires access to longer wavelength IR data. As mentioned before, in our sample, only very dusty star-forming galaxies with bright dust continua are detected in {\em Herschel}/PACS at 100 and 160\,{\um} bands. Therefore, we rely on stacks of images to obtain a sufficiently high {\it S/N} to calculate robust IR luminosities.

Figure~\ref{fig:pah_ir} shows {\ratio} stacks in bins of metallicity and {\o32}. We combine the two lowest metallicity bins and the two highest {\o32} bins in Figure~\ref{fig:pah_ism} and Table~\ref{tab:pah_ism} to gain higher {\it S/N} in the PACS stacks. The N2 stacks are adopted only for galaxies at $z>2$, because otherwise the highest N2 metallicity bin is dominated by lower redshift ($z\sim 1.5$) galaxies. In O3N2 and {\o32} plots, the median redshifts of all bins are similar and all above $z=2$. We note that there are only 8 galaxies at $z<2$ in the {\o32} sample, as [O{\sc II}] is not covered by the MOSFIRE {\it Y} filter at $z<1.6$ (see Figure~\ref{fig:pah_ism}c). Properties of {\ratio} stacks are listed in Table~\ref{tab:pah_ir}.

To place our analysis in the context of other studies, we compare our results with E11, \citet[][R12]{reddy12a}, and \citet[][W08]{wuyts08}.
E11 found that typical main-sequence galaxies between $z=0-2.5$ follow a Gaussian distribution of $L_{\rm IR}/L_{8}$ ($\nu L_{\nu}(8\mu{\rm m})\equiv L_8$) centered at $L_{\rm IR}/L_{8}=4$ ($\sigma=1.6$) with a tail of starburst galaxies with larger $L_{\rm IR}/L_{8}$ ratios. The median $L_{\rm IR}/L_{8}$ ratio of all galaxies in their sample was $4.9^{+2.9}_{-2.2}$. Furthermore, the R12 study found $L_{\rm IR}/L_{8}=7.7\pm 1.6$ for a sample of UV-selected galaxies at $1.5\leq z<2.6$ (with mean redshift of 2.08). R12 attributed their higher $L_{\rm IR}/L_{8}$ to redshift evolution and larger IR luminosity surface densities compared to those of the local galaxies.
For a consistent comparison of these ratios with our results, we converted $L_{\rm 8}/L_{\rm IR}$ ratios of R12 and E11 to {\ratio} ratios. 
The E11 and R12 samples have {\lir} $=5\times 10^{9}-3\times 10^{12}\,\lsun$ and $10^{10}-5\times 10^{12}\,\lsun$, respectively.
According to the CE01 templates, $\nu L_{\nu}$ at 7.7\,{\um} is higher than $\nu L_{\nu}$ at 8\,{\um} by 29\% for a template with {\lir} $=5\times 10^{9}\,\lsun$ and 12\% for {\lir} $=5\times 10^{12}\,\lsun$.
The average {\lir} in the R12 and E11 high-z samples is $\sim 2\times 10^{11}\,\lsun$, which corresponds to a $L_{7.7}/L_8$ ratio of 1.25. Therefore, we adopt a correction factor of 1.25 to convert the E11 and R12 $L_{\rm 8}/L_{\rm IR}$ ratios to {\ratio} ratios. The variations of the {\ratio} ratios resulted from adopting other conversion factors (from 1.12 to 1.29) are well within the reported 1\,$\sigma$ measurement uncertainty of {\ratio}.
After the conversion, we obtain {\ratio}$=0.25^{+0.15}_{-0.11}$ and $0.16\pm 0.03$ for the E11 and R12 studies, respectively. These ratios and their confidence intervals are shown with lines and shaded regions in Figure~\ref{fig:pah_ir}.
The {\ratio} of E11 is consistent with our highest metallicity and lowest {\o32} stacks. This is expected as at the same stellar mass, low-redshift galaxies have typically higher metallicities and less intense ionizing radiation fields. The {\ratio} of R12 is in agreement with our lower metallicity stacks and the middle {\o32} stack, as well as with the stacks of the full sample (yellow stars).

We also compare our results with those of \citet{wuyts08, wuyts11a}. In these studies, the authors derived an empirical IR template and luminosity-independent monochromatic conversions from 24\,{\um} flux density to {\lir} as a function of redshift. We used the median redshifts of our three {\o32} stacks (from the lowest to the highest bin: $z=2.29$, 2.27, and 2.29, Table~\ref{tab:pah_ir}) to adopt appropriate $f_{24}$-to-{\lir} conversions and compare them with our results in Figure~\ref{fig:pah_ir}(d). The width of the purple line indicates the range of the W08 $f_{24}/L_{\rm IR}$ ratios for different redshifts of our three stacks. As expected, $f_{24}$/{\lir} of W08 is consistent with our middle {\o32} bin as well as the stack of all objects, but there is discrepancy with the low and high {\o32} stacks. The same conclusion holds for the metallicity stacks. In Section~\ref{sec:implications}, we will discuss the implications of the inconsistency between the $L_{7.7}$-to-{\lir} conversions of W08, E11, and R12 with our observed values at low stellar masses.

\begin{figure}[tbp]
	\subfigure{
		\includegraphics[width=.49\textwidth]{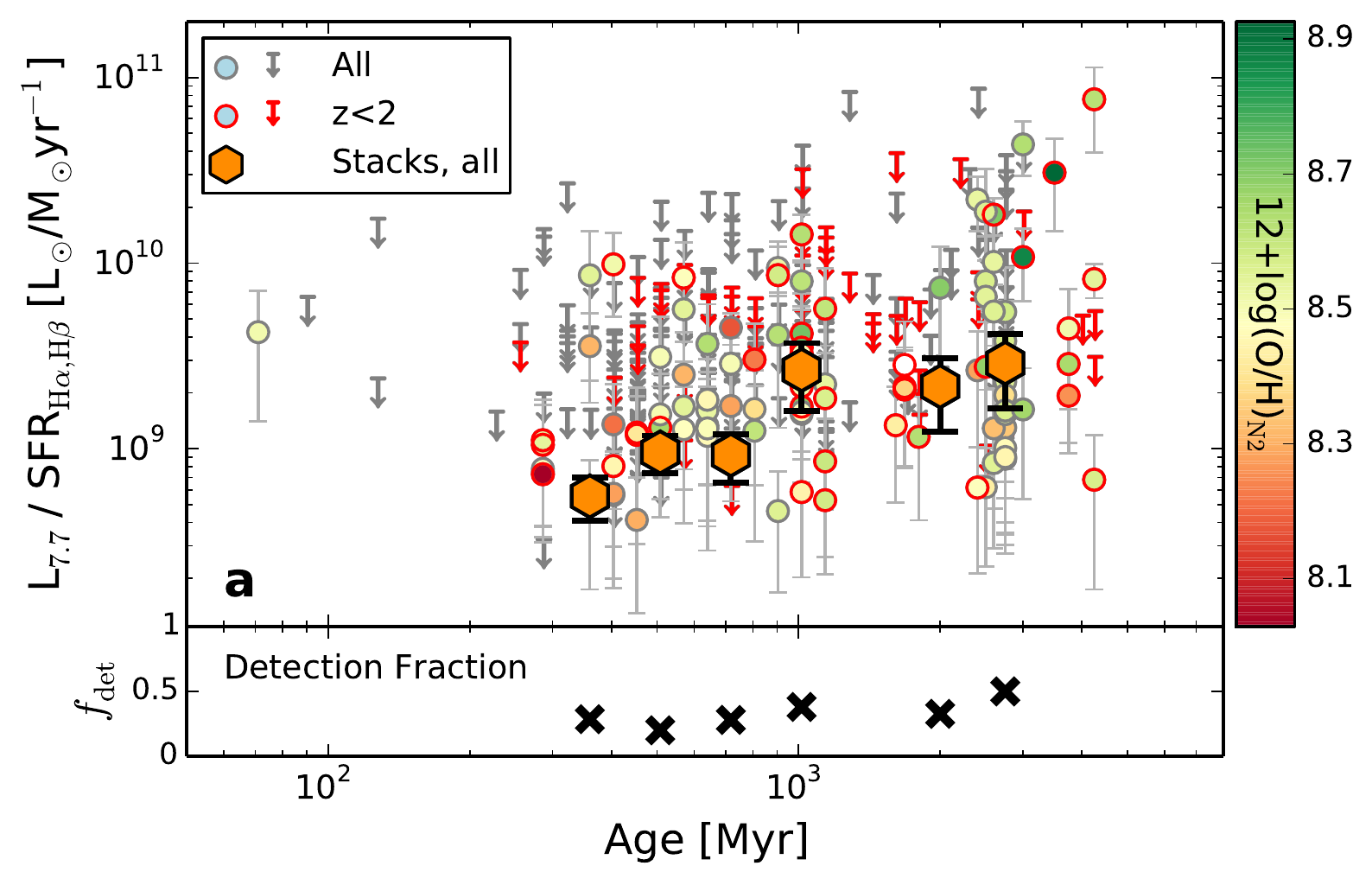}}
	\\
	\subfigure{
		\includegraphics[width=.35\textwidth]{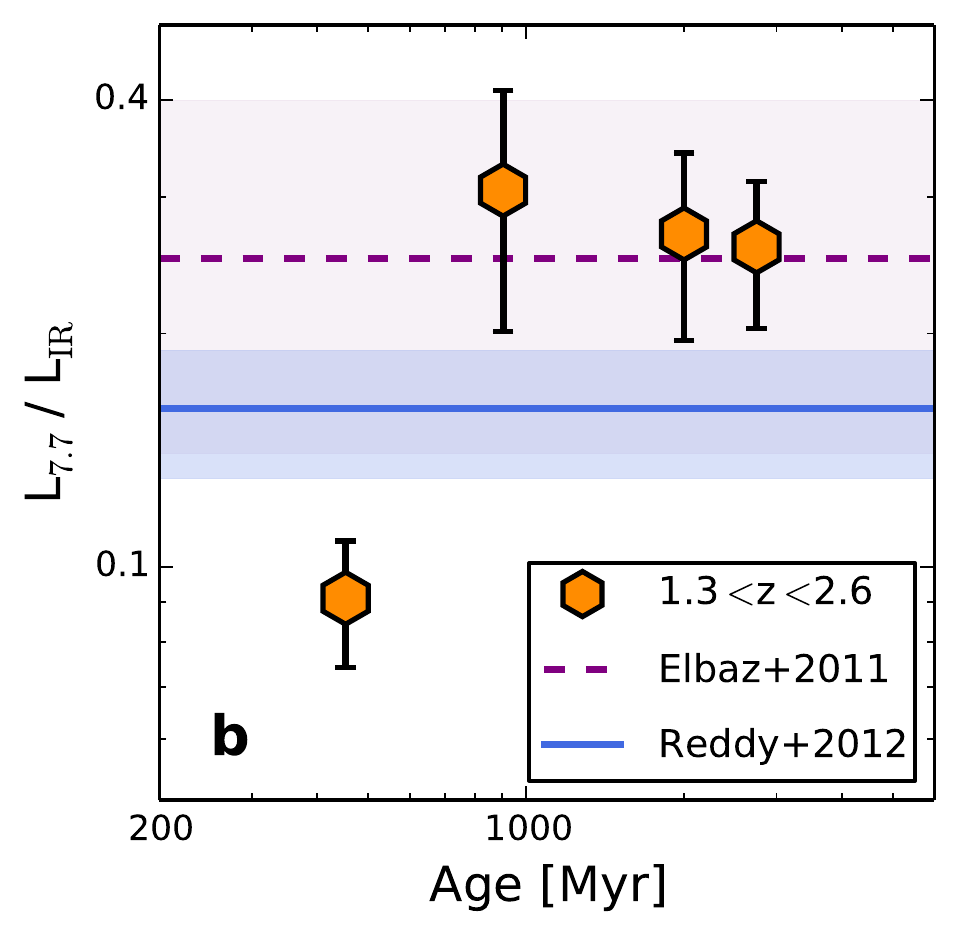}}
		\caption{(a): ratio of $L_{7.7}$ to {\sfr} as a function of age. Symbols are the same as Figure~\ref{fig:pah_ism}. Individual galaxies are color coded based on their N2 metallicity. (b): ratio of 7.7\,{\um} luminosity to {\lir} as a function of age. Symbols are the same as Figure~\ref{fig:pah_ir}. Ages are derived from the best-fit SEDs.
		Stacked values are listed in Tables~\ref{tab:pah_ism} and \ref{tab:pah_ir}.		
		}		
	\label{fig:pah_age}
\end{figure}

\section{PAH and Age}
\label{sec:pah_age}

PAH molecules are thought to form in the outflows of carbon-rich AGB stars \citep[e.g.,][]{latter91,tielens08}. These evolved stars begin enriching the ISM after their death (at an age of $\sim 400$\,Myr, \citealt{galliano11}), as opposed to core-collapse supernovae explosions, which enrich the ISM with dust on short timescales following the onset of star formation ($\lesssim 10$\,Myr). As a result, some studies suggest that the correlation between the PAH intensity and metallicity arises because of the delayed formation of the PAH molecules in chemically young systems \citep{dwek05,galliano08}. \citet{galliano08} modeled the dust production associated with massive and AGB stars and showed that PAHs are less abundant in young metal-poor systems. To test this scenario, we investigate the PAH intensity as a function of age in our sample.

In Figure~\ref{fig:pah_age}, we show the trend of $L_{7.7}$/SFR and {\ratio} as a function of age. Ages are derived from the best-fit SED models (Section~\ref{sec:data_sfr}). The PAH intensity is essentially constant within the errors at ages $\gtrsim 900$\,Myr, but both the $L_{7.7}$/SFR and {\ratio} ratios drop significantly by a factor of $\sim 3$ for young galaxies with ages $\sim 300-800$\,Myr. 
Galaxies with ages $\lesssim 500$\,Myr (the first bin in Figure~\ref{fig:pah_age}a) have a wide range of metallicities ($12+\log({\rm O/H})_{\rm O3N2}\sim 8-8.6$), although the metallicities are systematically lower (median of 8.2) compared to that of the whole sample (median of 8.3).
To demonstrate that metallicity can not completely account for the variation of $L_{7.7}$/{\sfr} with age, we randomly draw galaxies from the old population (with ages $> 500$\,Myr), such that the subsample has the same metallicity distribution as that of the young population (with ages $\leq 500$\,Myr). We stack the 24\,{\um} images, and find that the average $L_{7.7}$/SFR of the old population subsample is a factor 3 larger than the average $L_{7.7}$/SFR of the young population, although they have the same metallicity distribution.
This indicates that the observed deficit of PAH emission in young galaxies cannot be {\em fully explained} by low metallicity and high ionization parameter.

\begin{figure*}[tbp]
	\subfigure{
		\includegraphics[width=.42\textwidth]{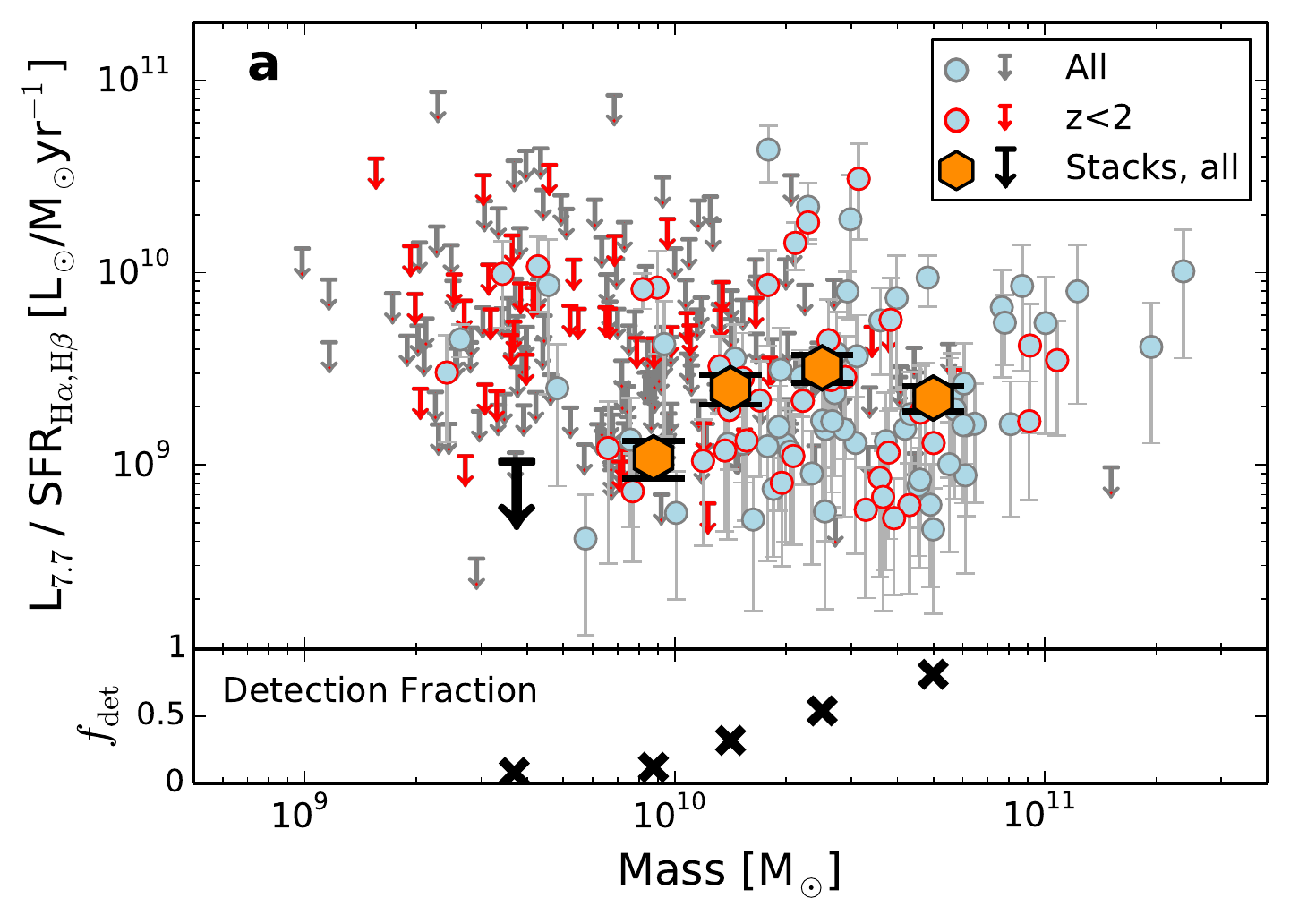}}
	\quad
	\subfigure{
		\includegraphics[width=.28\textwidth]{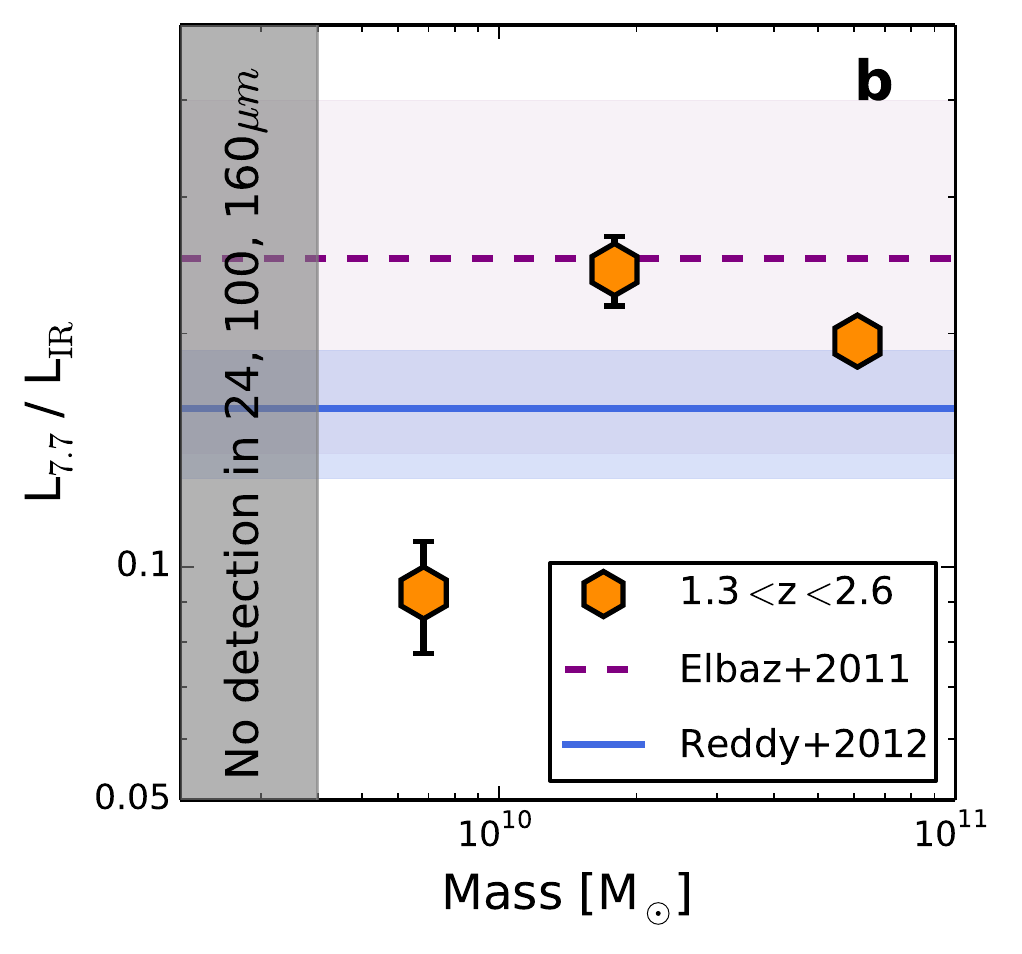}}
	\quad
	\subfigure{
		\includegraphics[width=.29\textwidth]{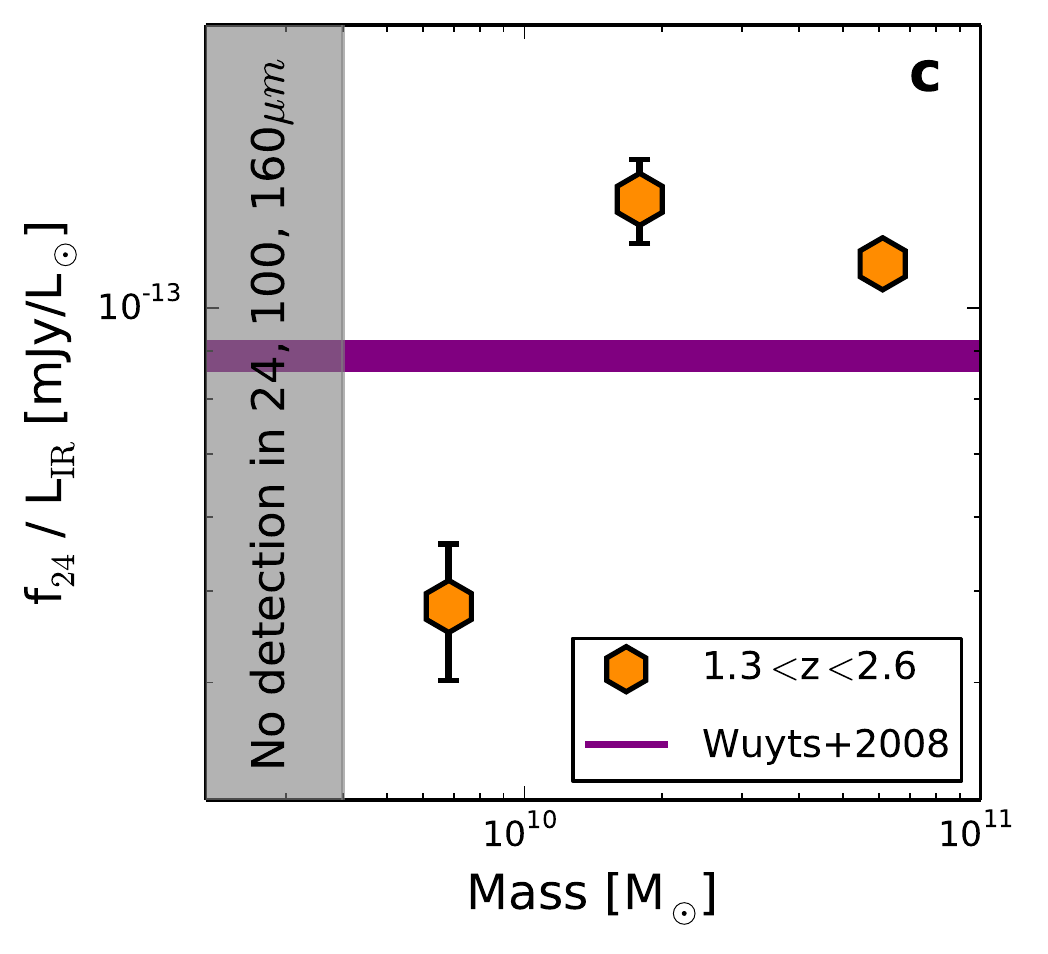}}
		\caption{(a): ratio of $L_{7.7}$ to {\sfr} as a function of stellar mass. Symbols are the same as Figure~\ref{fig:pah_ism}. (b) and (c): ratio of 7.7\,{\um} luminosity and 24\,{\um} flux density to {\lir} as a function of $M_*$, respectively. Symbols are the same as Figure~\ref{fig:pah_ir}. In the lowest mass bin ($M_*\leq 10^{9.6}\,\msun$) none of the 24\,{\um}, 100\,{\um}, and 160\,{\um} fluxes are detected at 2.5$\sigma$.
		Stacked values are listed in Tables~\ref{tab:pah_ism} and \ref{tab:pah_ir}.		
		}		
	\label{fig:pah_mass}
\end{figure*}

\section{L$_{7.7}$ as a tracer of Total IR Luminosity: Implications for High-\lowercase{z} studies}
\label{sec:implications}

As shown in Figures~\ref{fig:pah_ism} and \ref{fig:pah_ir} and discussed in Section~\ref{sec:pah_ism}, the ratio of rest-frame $L_{7.7}$ to SFR or {\lir} strongly depends on metallicity and {\o32}. Figure~\ref{fig:pah_mass} shows the correlation between {\ratio} and $M_*$, which likely arises from the correlation between $M_*$ and metallicity \citep{tremonti04} and the anti-correlation between $M_*$ and {\o32} \citep[e.g.,][]{sanders16a}.

Similar to the 24\,{\um}-to-IR conversions of W08, E11, and R12, luminosity-dependent conversions of \citet{caputi07}, \citet{bavouzet08}, \citet{rigby08}, and \citet{reddy10} are also only consistent with massive galaxies with $M_*\gtrsim 10^{10}\,\msun$ and $L_{\rm IR}\gtrsim 10^{11-11.5}\,\lsun$. In the next two sections we will discuss the implications of this result for high-redshift studies.

\begin{figure}[tbp]
	\includegraphics[width=.5\textwidth]{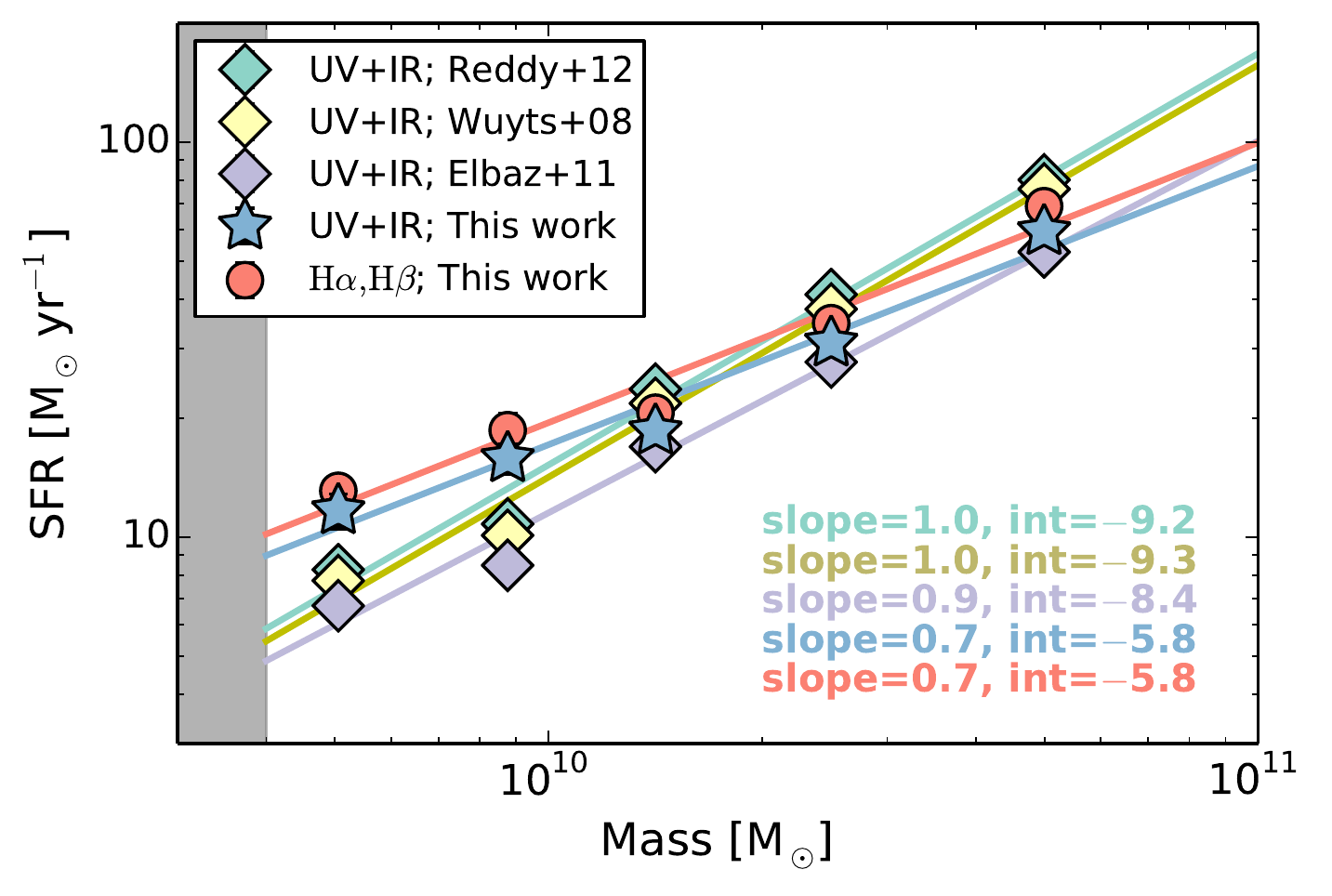}
	\caption{SFR in bins of stellar mass. Orange circles are 3$\sigma$-clipped means of dust-corrected {\sfr} in each bin. Other symbols show the sum of SFR$_{\rm UV}$ at 1600\AA~and SFR$_{\rm IR}$ derived from 24\,{\um} stacks using different conversions. Blue stars adopt the conversion derived in this study (Figure~\ref{fig:pah_mass}b): {\ratio}$=0.09$ and 0.22 for $M_*$ below and above $10^{10}\,\msun$, respectively. Cyan, yellow, and purple diamonds are from single-value conversions of R12, W08, and E11, respectively. We use the median redshift in each bin to adopt the corresponding W08 conversion. 
	The SFR$_{\rm IR}$ and {\sfr} errorbars are the same size as the symbols.
	The best-fit lines, slopes, and intercepts estimated from a simple linear least-squares regression to $\log({\rm SFR})$ vs. $\log(M_*)$ are shown on the plot with respective colors.
	The grey region shows $M*\leq 10^{9.6}\,\msun$, where the 24\,{\um} stack is not detected.
		}		
	\label{fig:sfr_mass}
\end{figure}

\subsection{The SFR-$M_*$ relation}
\label{sec:sfr-mass}

In \citet{shivaei15b} we discussed how SFR-$M_*$ studies at $z\sim2$ that rely on {\halpha} SFRs predict shallower log(SFR)-log($M_*$) slopes \citep[$\sim0.5-0.7$,][]{erb06c,zahid12,koyama13,atek14,darvish14,shivaei15b}, compared to those that utilize IR SFRs \citep[slope of $\sim 1$,][]{daddi07a,santini09,wuyts11b,reddy12b,whitaker14b}. In order to understand the cause of this discrepancy, in \citet{shivaei16a} we compared {\sfr} with SFRs inferred from panchromatic SED models that were fit to rest-frame UV to far-IR photometry. The comparison indicated that {\sfr} does not underpredict total SFR even for the most star-forming and dusty galaxies in our sample (up to $\sim 300\,\msun\,{\rm yr^{-1}}$). 
In this section, we examine whether the discrepancy in the slope of the SFR-$M_*$ relation arises from the 24\,{\um}-inferred IR SFRs that are not corrected for the PAH-metallicity effect.

Figure~\ref{fig:sfr_mass} shows SFR in bins of stellar mass for galaxies with $M_*\geq 10^{9.6}\,\msun$. In Figure~\ref{fig:sfr_mass}, orange circles denote mean {\sfr} and other symbols are mean SFR$_{\rm UV}$ added to mean SFR$_{\rm IR}$, where the latter is derived from various conversions of $f_{24}$ to {\lir} adopted from the literature (diamonds) and from this analysis (blue stars), which is described as follows.
We derive a mass-dependent conversion of $L_{7.7}$ to {\lir} based on our stacks presented in Figure~\ref{fig:pah_mass}(b,c) and Table~\ref{tab:pah_ir}. The {\ratio} ratio is $0.09, 0.24,$ and 0.20 for $\log(M_*/{\rm M_{\odot}})=9.6-10.0, 10.0-10.6$, and $10.6-11.6$, respectively. We adopt {\ratio} $=0.09$ for $M_* < 10^{10}\,\msun$ and an average of the two highest mass bins, 0.22, for $M_*\geq 10^{10}\,\msun$\footnote{We express our results in terms of mass dependence so as to avoid biases that may be introduced by excluding objects without detections of some of the strong lines required to calculate metallicities. The metallicity dependence of the {\ratio} ratios are reported in Table~\ref{tab:pah_ir}.}. The result is shown with blue stars in Figure~\ref{fig:sfr_mass}.
The diamonds in Figure~\ref{fig:sfr_mass} reflect results of the mass-independent 24\,{\um}-to-IR conversions of R12, W08, and E11. The 24\,{\um} stack below $10^{9.6}\,\msun$ (shaded grey region in Figure~\ref{fig:sfr_mass}) is not robustly detected, owing to the reduced dust emission at low masses and low SFRs.

The SFR$_{\rm UV}+{\rm SFR}_{\rm IR}$ derived from the mass-dependent 24\,{\um}-to-IR conversion presented in this study is in a very good agreement with {\sfr}, particularly considering that these two SFRs are derived completely independently from each other.
As expected, the other IR SFRs that are inferred from a single 24\,{\um}-to-IR conversion are underestimated for low mass ($M_* < 10^{10}\,\msun$) galaxies. Figure~\ref{fig:sfr_mass} clearly demonstrates how using a single 24\,{\um}-to-IR conversion can underestimate SFRs at lower masses and lead to a steeper log(SFR)-log($M_*$) slope compared to that derived in this study. A simple linear least-squares regression calculation gives a slope of $1.0\pm 0.1$ for the stacks of SFR$_{\rm UV}+{\rm SFR}_{\rm IR}$ that assume the W08, E11, and R12 conversions and a slope of $0.7\pm 0.1$ for the {\sfr} and the SFR$_{\rm UV}+{\rm SFR}_{\rm IR}$ stacks that adopt our mass-dependent conversion. 

The slope of 0.7 found in this study is consistent with observations at lower redshifts (slope $\sim 0.7$ at $z\sim 0.1$ and 1, \citealt{salim07} and \citealt{noeske07a}, respectively), suggesting no evolution in the slope of the SFR-$M_*$ relation from $z\sim 0$ to 2. Moreover, the increased SFR$_{\rm UV}+{\rm SFR}_{\rm IR}$ of the low-mass galaxies estimated from our mass-dependent 24\,{\um}-to-IR conversion, removes evidence for a ``flattening'' at the high-mass end of the SFR-$M_*$ relation, as suggested in some previous studies \citep[e.g.,][]{whitaker14b,schreiber15}.
However, it is plausible that the slope of the relation steepens at masses lower than those accessible in this study -- a mass-independent shallow slope of 0.7 would lead to a too rapid evolution in the stellar mass function at low masses \citep{weinmann12,leja15}.

Our results indicate that specific SFR (sSFR $\equiv$ SFR/$M_*$) is a decreasing function of $M_*$. If we assume that the gas and stars occupy the same regions in the galaxies, then the Schmidt relation \citep{kennicutt98a} implies that ${\rm SFR}\propto M_{\rm gas}^{1.4}$. Consequently, sSFR is related to the cold gas fraction defined as $M_{\rm gas}/(M_*+M_{\rm gas})$ \citep{reddy06b}. The negative slope of the log(sSFR)-log($M_*$) relation indicates that, as expected, there is a correspondence between the cold gas fraction and stellar mass.

There is tension between our results and those from simulations. According to simulations, the slope of the SFR-$M_*$ relation is close to unity and the sSFR is almost a constant function of mass \citep[][among many others]{dutton10a,dave11a,torrey14,kannan14,sparre15a}. 
Furthermore, our analysis shows an increased sSFR at $M_*=5\times 10^{9}$ at $z\sim 2$ by a factor of 1.8 ($\log({\rm sSFR/Gyr})=0.4$), compared to the previous sSFR measurements that used mass-independent 24{\um}-to-IR conversions. 
The increased sSFR found here also aggravates the inconsistency between observations and models, where the models predict lower sSFRs at $z\sim 2$ than those reported in previous studies \citep{furlong15,sparre15a,dave16}. From the observational viewpoint, we showed in \citet{shivaei15a} that our MOSDEF {\halpha}-detected sample is complete down to $M_*=10^{9.5}\,\msun$. Therefore, the shallow slope of 0.7 in this study is immune from the common sample selection biases at low masses. On the other hand, \citet{dave08} and \citet{torrey14} demonstrated that changing feedback models in simulations does not affect the slope of the relation, as it moves galaxies along the SFR-$M_*$ relation. More investigations are needed to resolve these issues and improve the agreement between simulations and observations.

\subsection{Bolometric luminosity density at $z\sim 2$}

Luminous infrared galaxies (LIRGs) with $L_{\rm IR}=10^{11}-10^{12}\,\lsun$, are known to contribute significantly to the total IR luminosity density at $z\sim 2$ \citep{reddy08,rodighiero10,magnelli11}. 
Current measurements of the IR luminosity density (and the obscured SFR density) at $z\sim 2$ are either based solely on ultra luminous infrared galaxies (ULIRGs) with $L_{\rm IR}>10^{12}\,\lsun$ \citep[e.g.,][]{gruppioni13} or have included lower IR luminosities by relying on different methods of converting 24\,{\um} flux density to {\lir} \citep[e.g.,][]{perez05,caputi07,reddy08,reddy10,rodighiero10,riguccini11}. The most common method uses empirical IR templates that are calibrated with local galaxies (e.g., CE01, \citealt{dh02}, \citealt{rieke09}). These IR templates are shown to overestimate {\lir} for ULIRGs at $z\sim 2$ \citep{nordon10,elbaz11,reddy12a}. To overcome this so-called ``mid-IR excess,'' other 24\,{\um}-to-IR conversions that were calibrated with samples of high-redshift galaxies were introduced \citep[][among many others]{bavouzet08,elbaz11}. However, the dependence of {\ratio} on metallicity and stellar mass was neglected in these calibrations. As we showed in previous sections, these conversions underpredict {\lir} at low and moderate masses ($M_*\lesssim 10^{10}\,\msun$) and IR luminosities ($L_{\rm IR}\lesssim 10^{11}\,\lsun$).
 
The increased {\lir} of the low and moderately IR luminous galaxies leads to a higher IR luminosity density at $z\sim 2$. To quantify this effect, we simulate 1000 galaxies drawn from the double power law luminosity function of \citet{magnelli11}, increase the IR luminosities of galaxies below $L_{\rm IR}=10^{11}\,\lsun$ in accordance with the relations found in this study (as discussed below), bin the galaxies (we used bins with 0.3\,dex width, similar to \citealt{magnelli11}), and re-fit the bins with the double power law function. We use two alternative methods to include the increased {\lir} of low-luminosity galaxies. In the first method, we increase {\lir} by a factor of 2 for all simulated galaxies with $L_{\rm IR}\leq 10^{11}\,\lsun$.
This assumption is motivated by our finding for the low-mass galaxies in the previous section (Section~\ref{sec:sfr-mass}). However, from our data it is not clear whether {\lir} of faint IR galaxies should be increased by a greater factor or an equal factor compared to that of the moderately IR luminous galaxies. Therefore, as a sanity check we adopt a second approach, where {\lir} is increased by a factor of 2 for galaxies with $L_{\rm IR}\leq 10^{10}\,\lsun$, and by a factor of 1.5 for those with $10^{10}<L_{\rm IR}\leq 10^{11}\,\lsun$.
In both cases, the number density of IR bins below $10^{11}\,\lsun$ increases by $\gtrsim 0.3$\,dex, and the faint-end slope steepens from $-0.6$ to $\sim -0.7$. The increase in the slope is within its $1\sigma$ error of 0.1 \citep{sanders03,magnelli11}. A stronger variation in {\lir} would cause a steeper faint-end slope, resulting in a higher fractional contribution of galaxies with $L_{\rm IR}\lesssim 10^{11}\lsun$ to the total IR budget and obscured SFR density at $z\sim 2$ than predicted before \citep{caputi07,reddy10}.
 
If we assume the same fractional contribution of the faint and moderately luminous IR galaxies to the bolometric (i.e., IR$+$UV) luminosity density at $z\sim 2$, as derived in previous studies \citep[$\sim 50-80\%$,][]{caputi07,reddy08,rodighiero10}, our estimate of a factor of $\sim 2$ increase in the IR luminosity of galaxies with $L_{\rm IR}\lesssim 10^{11}\lsun$ would increase the IR luminosity density by 30\%. Applying the 30\% increase to the IR luminosity density measurement of \citet{reddy08} at $z\sim 2$, and converting it to IR SFR density using the \citet{kennicutt98} relation, modified for a \citet{chabrier03} IMF\footnote{Assuming a \citet{salpeter55} IMF increases the SFR density values by a factor of $\sim 1.8$.}, yields an IR SFR density of 0.15\,${\msun}$\,yr$^{-1}$\,Mpc$^{-3}$. Adding this value to the UV SFR density measured in the same study \citep{reddy08} results in a total SFR density of 0.18\,${\msun}$\,yr$^{-1}$\,Mpc$^{-3}$ (assuming a \citet{chabrier03} IMF), which is $\sim 30\%$ higher than the initial value calculated from the unchanged IR luminosity density.
This finding exacerbates the discrepancy between the stellar mass density at $z\sim 2$ measured from observations and that obtained by integrating the SFR density over time \citep{hopkins06,madau14}.

\section{Why is PAH intensity correlated with metallicity?}
\label{sec:discussion}

The correlation between PAH intensity and metallicity has been observed and studied extensively in the local universe (\citealt{engelbracht05} and \citealt{madden06} were among the first). Several scenarios have been proposed to explain the observed trend, which involve the production mechanisms of PAH molecules or various ways of destroying them.

The origin of PAHs in galaxies is not completely understood, but evolved carbon stars are undoubtedly one of the main sources \citep{latter91,tielens08}. The presence of PAHs in the outflows of carbon-rich AGB stars is confirmed both in theoretical \citep{frenklach89,cau02,cherchneff06} and observational \citep{beintema96,boersma06} studies. As a consequence, it is expected that chemically young systems should have reduced PAH abundances \citep{dwek05,galliano08}. Also, PAH molecules are not as efficiently produced in low metallicity environments because fewer carbon atoms are available in the ISM.
On the other hand, PAHs can be destroyed in hostile environments such as supernovae shocks \citep{ohalloran06}. PAHs can also be effectively destroyed in low-metallicity environments with hard ionization fields, due to reduced shielding by dust grains \citep{madden06,smith07,hunt10}. Some authors have explored the size distribution of PAHs and suggested that there is a deficit of {\em small} PAH molecules at low metallicites \citep{hunt10,sandstrom10}.

The nature of the PAH-metallicity correlation is likely a combination of the production and destruction scenarios mentioned above. In our sample of galaxies at $z\sim 2$, we find strong trends between the PAH intensity and metallicity, {\o32}, and age, in agreement with both the formation and destruction scenarios. However, the significant increase in the fraction of undetected 24\,{\um} sources with increasing {\o32} (from 38\% to 91\%, Figure~\ref{fig:pah_ism}c) suggests that the destruction of PAHs in intense radiation fields may be the dominant physical mechanism. 
The low PAH emission could also be caused by the absorption of stellar UV photos by dust in the HII region before reaching the PAHs in the photodissociation region (PDR, \citealt{peeters04}). In this case, PAHs exist, but they are not exposed to the UV photons, and hence, do not emit in the mid-IR. However, in the case of a substantially dusty HII region, the recombination lines would also be systematically suppressed, thus resulting in systematic discrepancies between the UV- and {\halpha}-inferred SFRs. Such discrepancies are not observed in our sample \citep[see][]{reddy15,shivaei16a}.
Nevertheless, it is not trivial to completely differentiate between the various scenarios with our current data set.
 
A potentially important trend observed in Figure~\ref{fig:pah_mass} is that above $M_*\sim 10^{10}$ the {\ratio} (or $L_{7.7}$/SFR) reaches an almost constant value. This ``flattening'' is also seen in {\ratio} below ${\rm O}_{32}\sim 2.2$ (above $12+\log({\rm O/H)_{O_{32}}}\sim 8.2$, Figure~\ref{fig:pah_ism}c) and above the age of $\sim 900$\,Myr. 
We speculate that the observed metallicity (and mass) threshold reflects the point above which enough dust particles are produced to shield the ionizing photons and prevent the PAH destruction. This suggests that the constant value of {\ratio} at high metallicities is the ``equilibrium'' {\ratio} ratio. The suppressed {\ratio} at lower metallicities is indicative of preferential destruction of PAHs in environments with lower dust opacity, and hence, with harder and more intense radiation fields.
The paucity of PAH emission in young galaxies can also be attributed to the delayed production of PAHs by AGB stars and/or to the higher intensity of the radiation field in these young systems with high sSFR.
The threshold at $12+\log({\rm O/H)_{O_{32}}}\sim 8.2$ is the same as the value found by \citet{engelbracht05} and close to the value 8.1 found by \citet{draine07b} at $z\sim 0$.

\section{Summary}
\label{sec:conclusion}

We present the first results of investigating variations of the brightest PAH emission at 7.7{\um} with metallicity and ionizing radiation intensity at high redshift. For this study, we use the MOSDEF sample of star-forming galaxies at $1.37\leq z\leq 2.61$ that covers a broad range of metallicities ($\sim 0.2-1\,{\rm Z_{\odot}}$), stellar masses ($M_*\sim 10^9-10^{11.5}\msun$), and SFRs ($\sim 1-200\,\msun{\rm yr^{-1}}$). We adopt the N2 and O3N2 diagnostics of metallicity, and use {\o32} as a proxy for ionization parameter.
We quantify the PAH strength as the ratio of rest-frame 7.7{\um} luminosity ($\nu L_{\nu}(7.7\mu {\rm m})$), traced by {\em Spitzer}/MIPS 24{\um}, to dust-corrected {\sfr} and total IR luminosity. The main conclusions are as follows:

\begin{enumerate}

\item We find that, in agreement with studies of local galaxies, the PAH intensity depends strongly on gas-phase metallicity. The relative strength of $L_{7.7}$ to SFR decreases by a factor of $\sim 10$ from median metallicities of ${\rm Z}\sim 0.6$ to 0.3\,${\rm Z_{\odot}}$. 

\item There is a significant anti-correlation between the PAH intensity and {\o32}. This trend is stronger than the correlation of PAH with N2 and O3N2, as the majority of 24{\um} undetected galaxies have systematically high {\o32} ratios. The {\ratio} ratio changes from 29\% at {\o32} $\sim 0.9$ to 5\% at {\o32} $\sim 4$. The trends of the PAH strength with {\o32}, O3N2, and N2 suggest preferential destruction of PAH molecules in low metallicity environments, characterized with harder and more intense radiation fields.
Additionally, there is a sharp difference in the PAH intensity above and below {\o32} $\sim 2$ ($12+\log({\rm O/H)_{O_{32}}}\sim 8.2$). This threshold is also observed at $z\sim 0$. We speculate that above this metallicity threshold dust opacity is high enough that PAHs are no longer preferentially destroyed by ionizing photons, and hence, the {\ratio} ratio reaches a constant value.

\item For galaxies older than $\sim$1\,Gyr, we do not find a correlation between age and the PAH intensity. However, the {\ratio} ratios of the youngest quartile of galaxies in our sample (with ages $\sim 50-600$\,Myr) are significantly lower (by a factor of $\sim 3$) than those of galaxies with ages $\gtrsim 900$\,Myr. The low PAH intensity in young systems may be an indication of the delayed injection of PAH molecules to the ISM by carbon-rich AGB stars.

\item As a consequence of the mass-metallicity relation, we see a strong correlation between the PAH intensity and stellar mass. We show that commonly-used conversions of $L_8$ (or $f_{24}$) to {\lir} at $z\sim 2$ are only valid for massive and metal-rich galaxies. For galaxies with M$_* \lesssim 10^{10}\msun$, these conversions should be applied with caution as they underestimate the {\lir} and SFR by a factor of $\sim 2$.

\item The results of this analysis affect high-redshift studies that adopt mass (and metallicity) independent conversions of 24\,{\um} flux density to {\lir} and SFR over a large range of masses and metallicities. We show that by using our mass-dependent conversion of $L_{7.7}$ to {\lir} ({\ratio}$=$0.09 and 0.22 for $M_*$ below and above $10^{10}\,\msun$, respectively), the slope of the SFR$_{\rm UV+IR}$-$M_*$ relation decreases from unity to 0.7. The shallower slope of 0.7 is in agreement with the independently-derived slope of the {\sfr}-$M_*$ relation. Our results imply a higher sSFR (by a factor of 1.8) at $M_*\lesssim 10^{10}\msun$ compared to the previous IR$+$UV measurements. Based on this analysis, sSFR is a decreasing function of $M_*$, indicating a mass-dependent cold gas fraction in galaxies at $z\sim 2$. Our results are inconsistent with the unity slope of the SFR-$M_*$ relation derived from simulations, and add to the longstanding problem of models underestimating the sSFR at $z\sim 2$.

\item Our analysis suggests a higher bolometric luminosity density and SFR density by $\sim 30\%$ at $z\sim 2$, due to a factor of $\sim 2$ increase in the IR luminosity of low and moderately luminous IR galaxies ($L_{\rm IR}\lesssim 10^{11}\lsun$). This result reinstates the tension between the measured stellar mass density and the integral of SFR density at $z\sim 2$.

\end{enumerate}

We provide ratios of $L_{7.7}$-to-SFR and $L_{7.7}$-to-{\lir} as a function of metallicity, {\o32}, stellar mass, and age in Tables~\ref{tab:pah_ism} and \ref{tab:pah_ir} as a reference to be used for future studies of galaxies at $z\sim 2$.

In the future, with a larger sample of galaxies at $1.37\leq z\leq 2.0$, we will study the PAH intensity at different redshifts from $z\sim 1.5$ to 2.5, as well as incorporating $z\sim 0$ studies to construct a comprehensive picture of the evolution of PAHs over cosmic time. Moreover, future generations of ground- (e.g., TMT and E-ELT) and space-based (e.g., {\em JWST}) telescopes will significantly improve our knowledge about PAH molecules in the distant universe by extending the redshift range over which we can observe PAH emission and enabling us to probe a larger dynamic range of metallicity, {\o32}, age, and stellar mass.

\vspace{5mm}
The authors thank the referee for thoughtful suggestions.
We thank Mark Dickinson and Hanae Inami for providing part of the IR data, and Romeel Dav\'{e}, Lee Armus, George Rieke, David Cook, Marjin Franx, Joel Leja, and Louis Abramson for useful discussion. 
Support for IS is provided through the National Science Foundation Graduate Research Fellowship DGE-1326120. 
NAR is supported by an Alfred P. Sloan Research Fellowship.
Funding for the MOSDEF survey is provided by NSF AAG grants AST-1312780, 1312547, 1312764, and 1313171 and archival grant AR-13907, provided by NASA through a grant from the Space Telescope Science Institute.
The data presented herein were obtained at the W.M. Keck Observatory, which is operated as a scientific partnership among the California Institute of Technology, the University of California and the National Aeronautics and Space Administration. The Observatory was made possible by the generous financial support of the W.M. Keck Foundation.
The authors wish to recognize and acknowledge the very significant cultural role and reverence that the summit of Mauna Kea has always had within the indigenous Hawaiian community. We are most fortunate to have the opportunity to conduct observations from this mountain.

\bibliographystyle{apj}

\newpage

\begin{appendices}
\section{Ratio of Averages versus Average of Ratios}
\label{sec:appA}

Throughout the paper, we stack 24\,{\um} images and calculate average {\sfr} in bins of metallicity, {\o32}, mass, and age. It is important to note that an average of ratios, i.e. $\langle \frac{L_{7.7}}{{\rm SFR_{H\alpha,H\beta}}} \rangle$, and a ratio of averages, i.e. $\frac{\langle L_{7.7} \rangle}{\langle {\rm SFR_{H\alpha,H\beta}}\rangle}$, are not necessarily equal \citep[e.g.,][]{brown11}. 
Our method to compute the average of ratios is as follows.

First, we construct an inverse-{\sfr}-weighted average of 24{\um} images:
\begin{eqnarray}
\langle f_{24}\rangle_{\rm w} = \frac{\sum_i \frac{f_{24,i}}{\psi_i}}{\sum_i \frac{1}{\psi_i}}~,
\end{eqnarray}
where $\psi$ is {\sfr}. 
We need to measure $\langle f_{24}\rangle_{\rm w}$ so that we can fit 24\,{\um} flux densities to the CE01 IR templates and extract $\langle L_{7.7}\rangle_{\rm w}$, as follows. We calculate an inverse-{\sfr}-weighted average of redshifts (i.e., $\frac{\sum_i z_i/\psi_i}{\sum_i 1/\psi_i}$), shift the IR templates to the weighted-average redshift, and find the best-fit template through a least-$\chi^2$ method. $\langle L_{7.7}\rangle_{\rm w}$ is extracted from the best-fit model (see Section~\ref{sec:data_ir}).
Next, we calculate the weighted average of {\sfr}:
\begin{eqnarray}
\langle \psi \rangle_{\rm w} = \frac{\sum_i \frac{\psi_i}{\psi_i}}{\sum_i \frac{1}{\psi_i}} = \frac{N}{\sum_i \frac{1}{\psi_i}}~,
\end{eqnarray}
where $N$ is the total number of galaxies contributing to the stack. The ratio of $\langle L_{7.7}\rangle_{\rm w}$ to $\langle \psi \rangle_{\rm w}$ is mathematically equivalent to calculating the average of $L_{7.7}/\psi$ ratios ($\langle {L_{7.7}}/{\psi} \rangle$):
\begin{eqnarray}
\frac{\langle L_{7.7}\rangle_{\rm w}}{\langle \psi \rangle_{\rm w}} = \frac{\sum_i \frac{L_{7.7,i}}{\psi_i}}{\sum_i \frac{1}{\psi_i}} \times \frac{\sum_i \frac{1}{\psi_i}}{N}  
= \frac{\sum_i \frac{L_{7.7,i}}{\psi_i}}{N} 
= \Bigl \langle \frac{L_{7.7}}{\psi} \Bigl \rangle~.
\label{eq:3}
\end{eqnarray}

\begin{figure}
	\subfigure{
	\centering
		\includegraphics[width=.48\columnwidth]{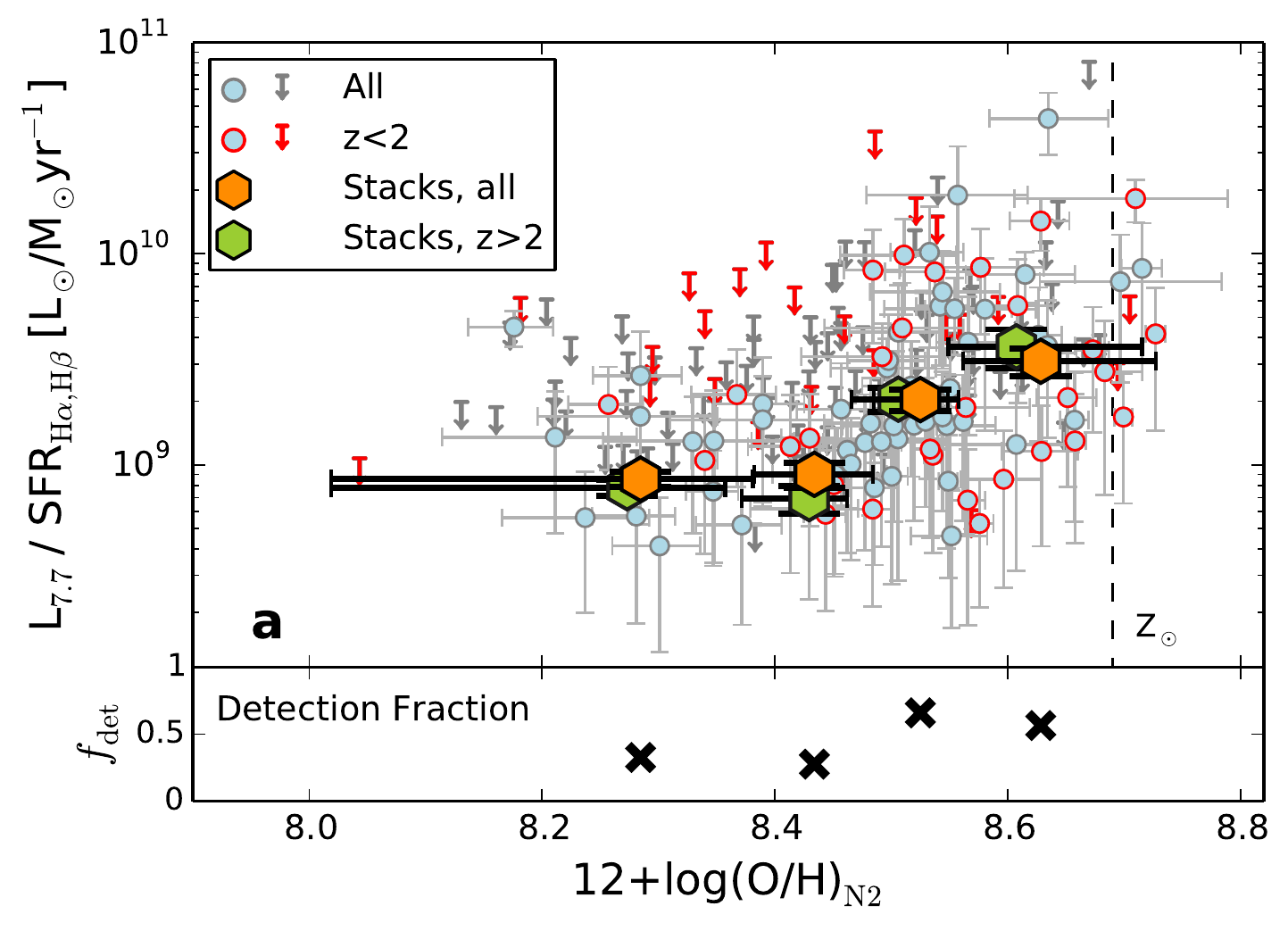}}
	\quad
	\subfigure{
	\centering
		\includegraphics[width=.48\columnwidth]{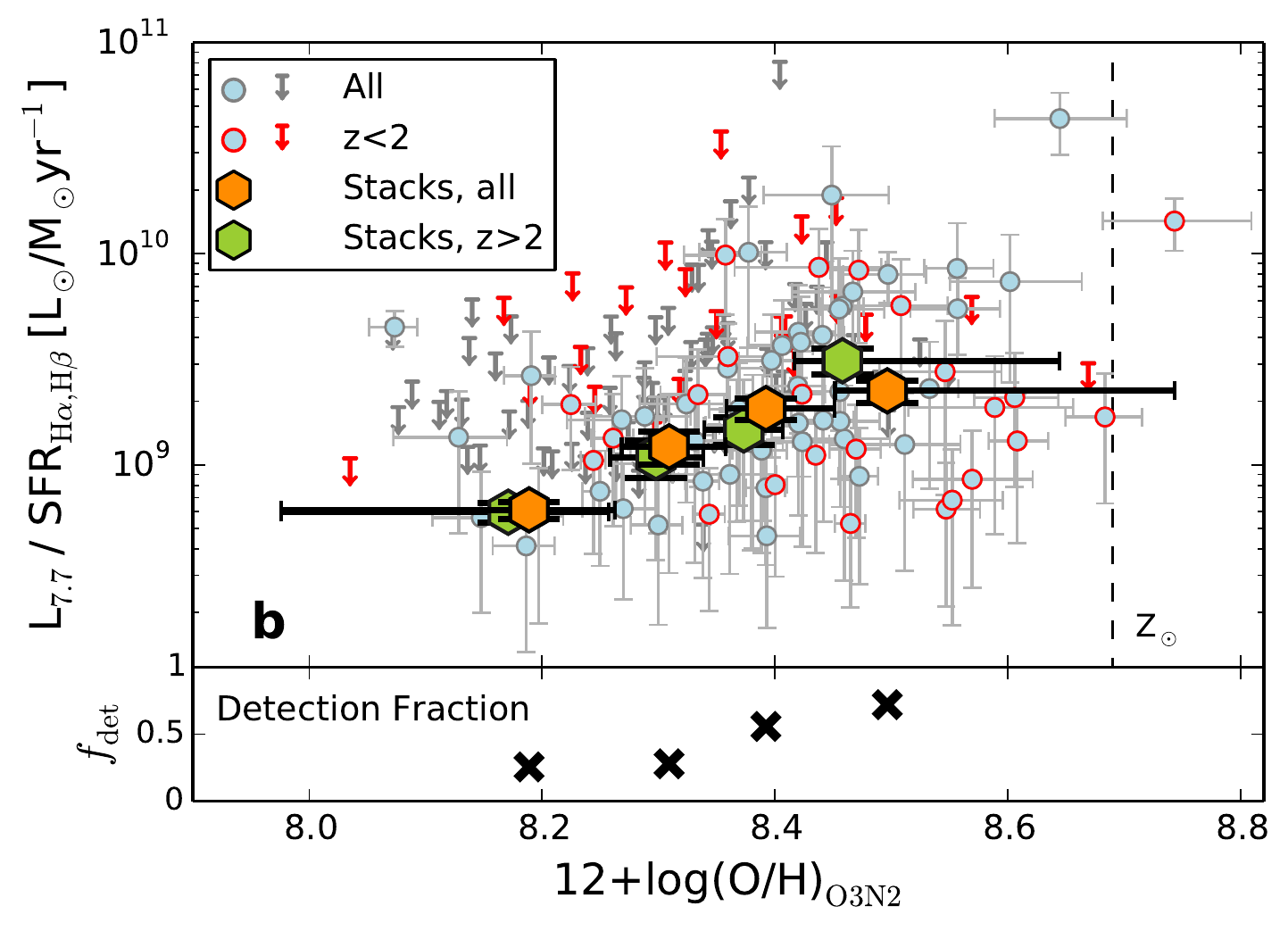}}
	\\
	\subfigure{
	\centering
		\includegraphics[width=.48\columnwidth]{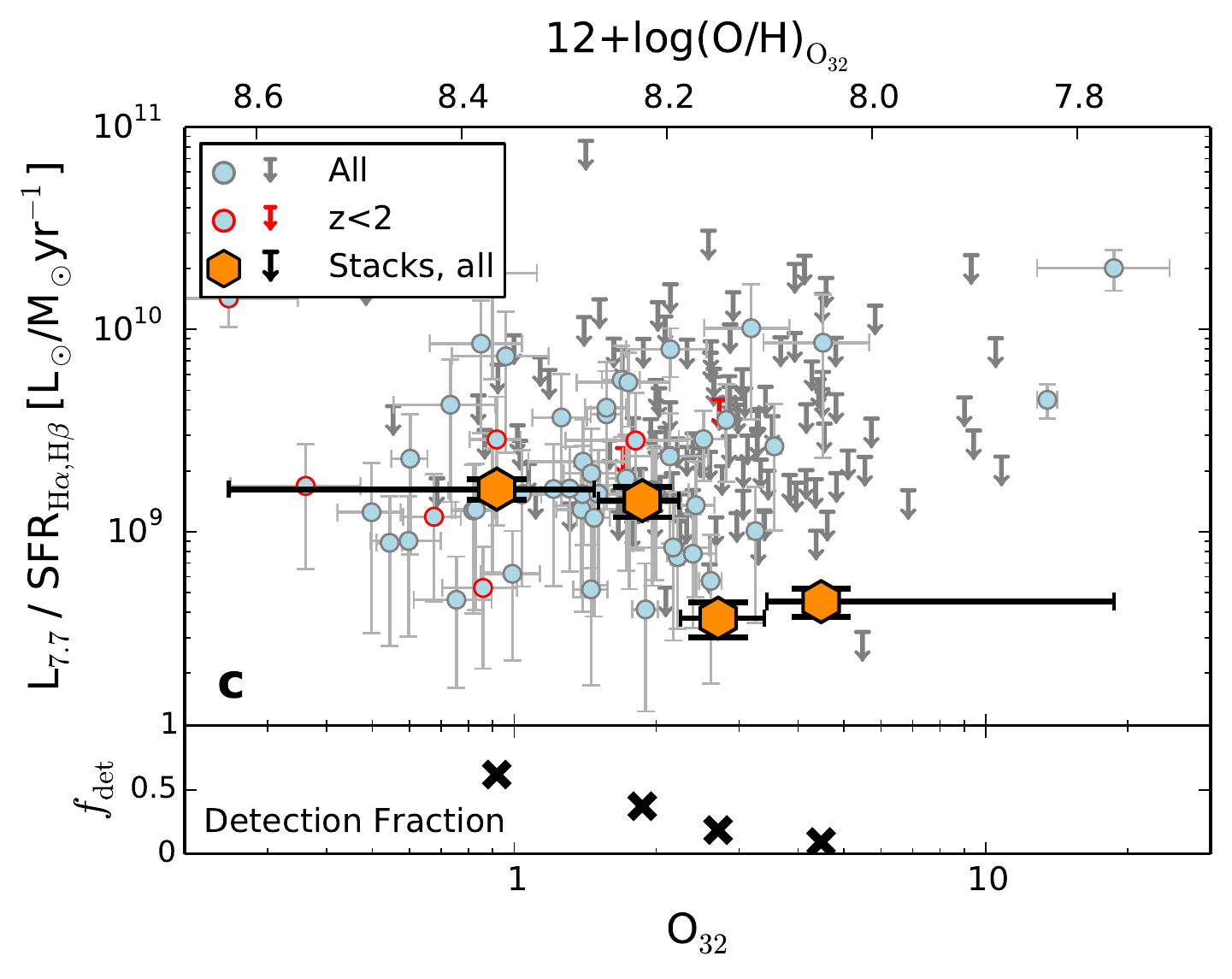}}
	\quad
	\subfigure{
	\centering
		\includegraphics[width=.48\textwidth]{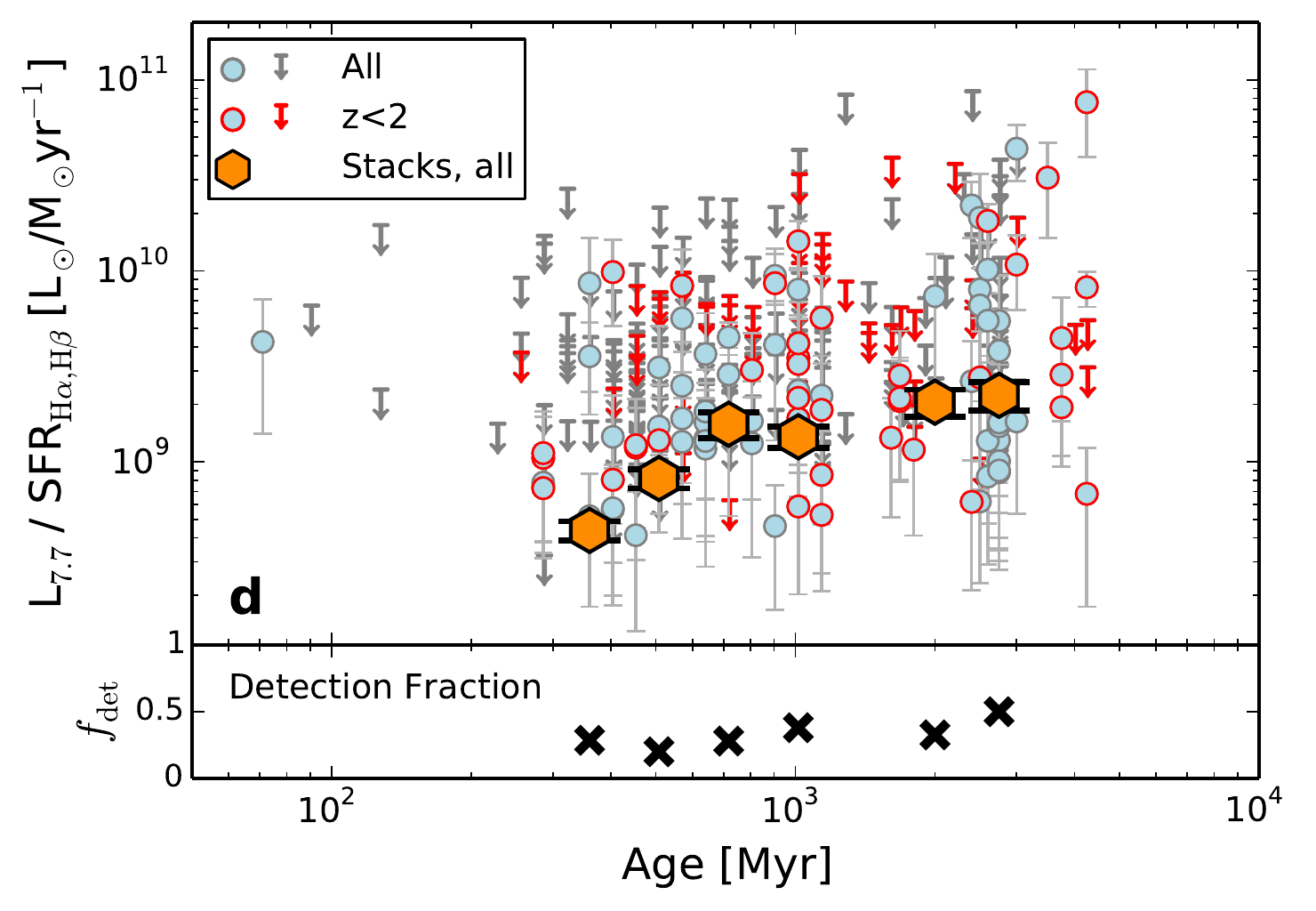}}
	\\
	\subfigure{
	\centering
		\includegraphics[width=.48\textwidth]{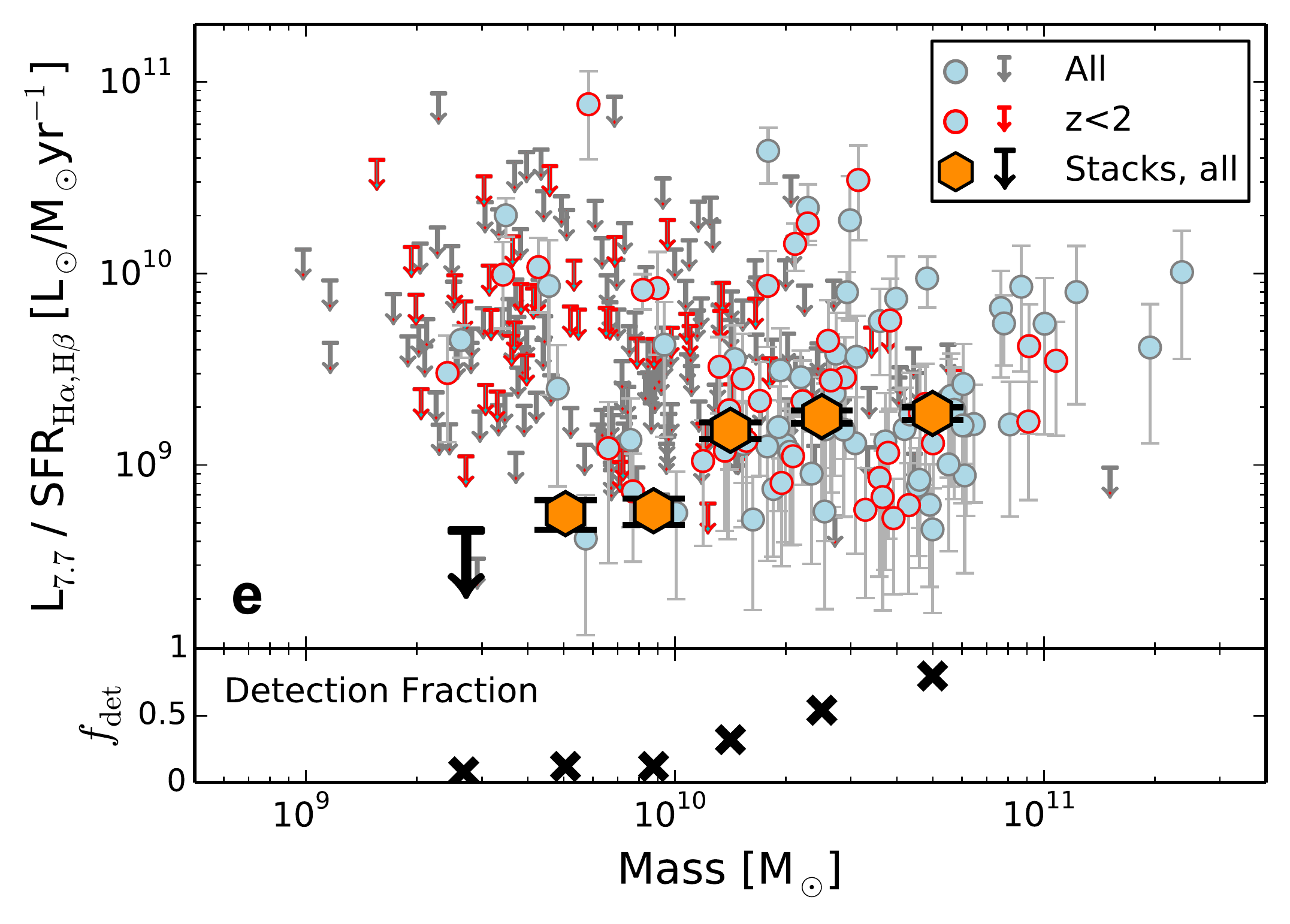}}

	\caption{Ratio of $L_{7.7}$ to {\sfr} for individual galaxies (small circles and arrows) and for average stacks of $L_{7.7}$ to 3$\sigma$-clipped averages of {\sfr} (hexagons and arrows) as a function of (a) N2 metallicity, (b) O3N2 metallicity, (c) {\o32}, (d) age, and (e) mass. Note that in these plots the orange symbols are ratios of average quantities (refer to Appendix~\ref{sec:appA}). The stacks and averages are in bins of the quantity on the horizontal axis.}
	\label{fig:ratio_avg}
\end{figure}

The $L_{7.7}$/{\sfr} plots throughout the paper (e.g., Figure~\ref{fig:pah_ism}) and the corresponding values in Table~\ref{tab:pah_ism} are derived according to Equation~\ref{eq:3}. In Figure~\ref{fig:ratio_avg}, we repeat the analyses using the ratios of averages, ${\langle L_{7.7} \rangle}/{\langle {\rm SFR_{H\alpha,H\beta}}\rangle}$, by simply dividing the $L_{7.7}$ average stacks by the 3$\sigma$-clipped averages of {\sfr}. For our sample, the ratio of averages and the average of ratios yield very similar results. The trends between the PAH intensity with metallicity, {\o32}, mass, and age are present regardless of the method adopted. Values of the $L_{7.7}$ stacks and the {\sfr} 3$\sigma$-clipped averages are also reported in Table~\ref{tab:pah_ism}.

As we do not have individual detections of objects in the PACS images, we cannot directly measure individual IR luminosities. As a result, we adopt a ratio of averages, i.e. ${\langle L_{7.7}\rangle}/{\langle L_{\rm IR}\rangle}$. As we demonstrated above with {\sfr}, adopting this method is likely to yield similar results to computing the average of ratios.
If anything, we speculate that $\langle {L_{7.7}}/{L_{\rm IR}} \rangle$ would result in a more significant trend between the PAH intensity and metallicity, compared to that of the ${\langle L_{7.7} \rangle}/{\langle L_{\rm IR}\rangle}$. This speculation is based on the fact that in the weighted average method, the low (or undetected) 24\,{\um} images are up-weighted, as they tend to have low {\sfr} (high $1/\psi$), and hence, in low metallicity bins the average of $L_{7.7}/L_{\rm IR}$ ratios would be even lower than the ratio of averages.

\end{appendices}
\newpage
\capstartfalse   
\begin{deluxetable*}{cccccccc}
\setlength{\tabcolsep}{0.05in} 
\tabletypesize{\footnotesize} 
\tablewidth{0pc}
\tablecaption{Properties of $L_{7.7}$/{\sfr} stacks}
\tablehead{
\colhead{Parameter} &
\colhead{Parameter Range} &
\colhead{$N$} &
\colhead{$\tilde{z}$} &
\colhead{$\langle f_{24}\rangle$ [${\rm \mu}$Jy]} &
\colhead{$\langle L_{7.7}\rangle$ [$10^9\,\lsun$]} &
\colhead{$\langle${\sfr}$\rangle$ [$\msun\,{\rm yr}^{-1}$]} &
\colhead{$\langle L_{7.7}$/{\sfr}$\rangle$ [$10^8\,\lsun/\msun\,{\rm yr}^{-1}$]}
}
\startdata
{$12+\log({\rm O/H})_{\rm N2}$} & {$8.04-8.37$} & {45} & {2.22} & {$16.4\pm 0.8$} & {$31.9\pm 1.5$} & {$37\pm 3$} & {$6\pm 1$} \\
{($N_{\rm tot}=185$)} & {$8.38-8.48$} & {47} & {2.23} & {$13.8\pm 0.8$} & {$27.2\pm 1.5$} & {$30\pm 4$} & {$5\pm 2$} \\
{} & {$8.48-8.56$} & {47} & {2.21} & {$39.9\pm 0.9$} & {$74.6\pm 1.6$} & {$37\pm 4$} & {$29\pm 11$} \\
{} & {$8.56-8.73$} & {46} & {2.17} & {$52.0\pm 0.9$} & {$90.2\pm 1.5$} & {$29\pm 4$} & {$41\pm 15$} \\ \hline \\

{$12+\log({\rm O/H})_{\rm O3N2}$} & {$8.03-8.26$} & {42} & {2.22} & {$10.8\pm 0.7$} & {$21.2\pm 1.3$} & {$35\pm 2$} & {$2.9\pm 0.7$} \\
{($N_{\rm tot}=171$)} & {$8.27-8.35$} & {43} & {2.21} & {$18.8\pm 0.8$} & {$35.9\pm 1.5$} & {$30\pm 5$} & {$7\pm 3$}\\
{} & {$8.35-8.44$} & {43} & {2.26} & {$28.4\pm 1.0$} & {$57.5\pm 2.1$} & {$31\pm 4$} & {$30\pm 9$} \\
{} & {$8.44-8.74$} & {43} & {2.13} & {$52.0\pm 1.0$} & {$86.4\pm 1.6$} & {$39\pm 5$} & {$35\pm 12$} \\ \hline \\

{{\o32}} & {$0.23-1.46$} & {42} & {2.29} & {$34.4\pm 1.1$} & {$72.7\pm 2.3$} & {$45\pm 5$} & {$26\pm 8$} \\
{($N_{\rm tot}=170$)} & {$1.46-2.21$} & {43} & {2.29} & {$17.3\pm 0.7$} & {$36.6\pm 1.5$} & {$26\pm 4$} & {$11\pm 3$} \\
{} & {$2.22-3.28$} & {43} & {2.27} & {$4.7\pm 0.7$} & {$9.9\pm 1.5$} & {$26\pm 3$} & {$4.8$}\tablenotemark{*} \\
{} & {$3.28-13.52$} & {42} & {2.29} & {$4.1\pm 0.6$} & {$8.9\pm 1.2$} & {$20\pm 2$} & {$4.4$}\tablenotemark{*} \\ \hline \\

{$\log(M_*/\msun$)} & {$9.00-9.84$} & {98} & {2.22} & {$0.9\pm 0.4$} & {$2.5$}\tablenotemark{*} & {$9\pm 1$} & {$7$}\tablenotemark{*} \\
{($N_{\rm tot}=296$)} & {$9.84-10.04$} & {50} & {2.29} & {$4.9\pm 0.6$} & {$10.8\pm 1$} & {$19\pm 2$} & {$11\pm 2$} \\
{} & {$10.04-10.28$} & {50} & {2.23} & {$16.0\pm 0.7$} & {$10.8\pm 1.3$} & {$31.3\pm 1$} & {$25\pm 4$} \\
{} & {$10.29-10.52$} & {50} & {2.27} & {$30.3\pm 0.8$} & {$31.3\pm 1.3$} & {$62.2\pm 2$} & {$32\pm 5$} \\
{} & {$10.53-11.37$} & {48} & {2.24} & {$69.5\pm 1.0$} & {$62.2\pm 1.8$} & {$130.9\pm 2$} & {$22\pm 3$} \\ \hline \\

{Age [Myr]} & {$71-453$} & {49} & {2.27} & {$6.5\pm 0.5$} & {$13.8\pm 1.1$} & {$30\pm 3$} & {$6\pm 1$}\\
{($N_{\rm tot}=296$)} & {$453-570$} & {50} & {2.23} & {$9.9\pm 0.6$} & {$19.4\pm 1.3$} & {$24\pm 2$} & {$10\pm 2$}\\
{} & {$570-904$} & {50} & {2.29} & {$12.6\pm 0.5$} & {$26.9\pm 1.1$} & {$17\pm 2$} & {$9\pm 3$}\\
{} & {$904-1278$} & {50} & {2.16} & {$18.1\pm 0.8$} & {$32.4\pm 1.4$} & {$24\pm 3$} & {$27\pm 11$}\\
{} & {$1278-2600$} & {49} & {2.18} & {$22.3\pm 1.0$} & {$41.0\pm 1.8$} & {$20\pm 3$} & {$22\pm 9$}\\
{} & {$2600-4250$} & {48} & {2.24} & {$34.9\pm 1.3$} & {$67.9\pm 2.5$} & {$29\pm 5$} & {$29\pm 12$}\\

\enddata

\tablenotetext{}{Notes. Entries show properties of the stacks shown with orange symbols ($1.37\leq z\leq 2.60$) in Figures~\ref{fig:pah_ism} and \ref{fig:pah_mass}. Each entry includes the parameter range, number of objects in each stack, median redshift, 24{\um} stacked flux density and its measurement uncertainty, rest-frame 7.7{\um} luminosity and its measurement uncertainty, 3$\sigma$-clipped mean {\sfr} and its error of the mean, and the average of $L_{7.7}$/{\sfr} ratios and its error. The latter should not be confused with the ratio of averages (refer to the text and Appendix~\ref{sec:appA}).}
\tablenotetext{*}{These object are undetected in the 24{\um} image stacks (before the aperture correction) and the values are 3$\sigma$ upper limits.}
\label{tab:pah_ism}
\end{deluxetable*}
\capstarttrue  

\capstartfalse   
\begin{deluxetable*}{cccccccccc}
	\setlength{\tabcolsep}{0.05in} 
	\tabletypesize{\footnotesize} 
	\tablewidth{0pc}
	\tablecaption{Properties of {\ratio} stacks}
	\tablehead{
	\colhead{Parameter} &
	\colhead{Parameter Range} &
	\colhead{$N$} &
	\colhead{$\tilde{z}$} &
	\colhead{$\langle f_{24} \rangle$ [${\rm \mu}$Jy]} &
	\colhead{$\langle f_{100} \rangle$ [${\rm \mu}$Jy]} &
	\colhead{$\langle f_{160} \rangle$ [${\rm \mu}$Jy]} &
	\colhead{$\langle L_{7.7} \rangle$ [$10^9\,\lsun$]} &
	\colhead{$\langle${\lir}$\rangle$ [$10^{10}\,\lsun$]\tablenotemark{a}} &
	\colhead{$\langle L_{7.7}\rangle$/$\langle${\lir}$\rangle$}
}
	\startdata
{$12+\log({\rm O/H})_{\rm N2}$}\tablenotemark{b} & {$8.02-8.47$} & {80} & {2.27} & {$12.4/pm 0.5$} & {$59\pm 30$} & {$472\pm 88$} & {$26\pm 1$} & {$12\pm 3$} & {$0.22\pm 0.05$}\\
{($N_{\rm tot}=160$)} & {$8.47-8.55$} & {40} & {2.27} & {$27.0\pm 0.9$} & {$347\pm 41$} & {$413\pm 126$} & {$56\pm 2$} & {$25\pm 4$} & {$0.22\pm 0.03$} \\
{} & {$8.55-8.72$} & {40} & {2.30} & {$34.0\pm 0.9$} & {$411\pm 51$} & {$486\pm 108$} & {$72\pm 2$} & {$26\pm 4$} & {$0.28\pm 0.04$} \\ \hline \\

{$12+\log({\rm O/H})_{\rm O3N2}$} & {$7.98-8.35$} & {86} & {2.21} & {$11.5\pm 0.4$} & {$199\pm 30$} & {$374\pm 98$} & {$22.3\pm 0.9$} & {$16\pm 3$} & {$0.14\pm 0.03$} \\
{($N_{\rm tot}=172$)} & {$8.35-8.44$} & {43} & {2.26} & {$26.6\pm 0.8$} & {$550\pm 50$} & {$454\pm 109$} & {$54\pm 2$} & {$30\pm 3$} & {$0.18\pm 0.02$} \\
{} & {$8.44-8.74$} & {43} & {2.13} & {$46.3\pm 0.9$} & {$616\pm 54$} & {$875\pm 110$} & {$77\pm 2$} & {$33\pm 3$} & {$0.24\pm 0.02$} \\ \hline \\

{{\o32}} & {$0.25-1.46$} & {42} & {2.29} & {$37.6\pm 0.8$} & {$329\pm 42$} & {$607\pm 105$} & {$79\pm 2$} & {$27\pm 3$} & {$0.29\pm 0.04$} \\
{($N_{\rm tot}=171$)} & {$1.46-2.71$} & {65} & {2.27} & {$10.8\pm 0.5$} & {$52\pm 36$} & {$441\pm 101$} & {$23\pm 1$} & {$11\pm 3$} & {$0.21\pm 0.06$} \\
{} & {$2.71-18.71$} & {64} & {2.29} & {$3.7\pm 0.4$} & {$286\pm 40$} & {$209\pm 81$} & {$8\pm 1$} & {$15\pm 2$} & {$0.05\pm 0.01$} \\ \hline \\

{$\log(M_*/\msun)$} & {$8.26-9.6$} & {92} & {2.15} & {$-3.2\pm 0.4$} & {$-125\pm 27$} & {$-281\pm 101$} & {$2.2$}\tablenotemark{*} & {$8.9$}\tablenotemark{*} & {--} \\
{($N_{\rm tot}=476$)} & {$9.6-10$} & {137} & {2.20} & {$4.7\pm 0.4$} & {$127\pm 27$} & {$247\pm 61$} & {$9.0\pm 0.7$} & {$10\pm 1$} & {$0.09\pm 0.02$}\\
{} & {$10-10.6$} & {173} & {2.19} & {$21.3\pm 0.5$} & {$261\pm 22$} & {$306\pm 63$} & {$39.5\pm 0.9$} & {$16\pm 2$} & {$0.24\pm 0.02$}\\
{} & {$10.6-11.6$} & {74} & {2.19} & {$69.4\pm 0.9$} & {$854\pm 38$} & {$1625\pm 102$} & {$121.7\pm 1.6$} & {$62\pm 3$} & {$0.20\pm 0.01$} \\ \hline \\

{Age [Myr]} & {$71-570$} & {100} & {2.26} & {$7.6\pm 0.4$} & {$172\pm 31$} & {$515\pm 84$} & {$15.8\pm 0.9$} & {$17\pm 3$} & {$0.09\pm 0.02$} \\
{($N_{\rm tot}=299$)} & {$570-1278$} & {100} & {2.22} & {$15.8\pm 0.5$} & {$291\pm 36$} & {$-166\pm 97$} & {$30.8\pm 1.0$} & {$10\pm 3$} & {$0.31\pm 0.11$}\\
{} & {$1278-2600$} & {50} & {2.18} & {$22.3\pm 1.0$} & {$318\pm 44$} & {$-88\pm 148$} & {$41.0\pm 1.8$} & {$15\pm 4$} & {$0.27\pm 0.07$}\\
{} & {$2600-4250$} & {49} & {2.23} & {$34.3\pm 1.3$} & {$263\pm 55$} & {$793\pm 131$} & {$65.8\pm 2.5$} & {$25\pm 5$} & {$0.26\pm 0.06$}
	\enddata

	\tablenotetext{}{Notes. Entries show properties of stacks in Figures~\ref{fig:pah_ir} and \ref{fig:pah_mass}. Each entry includes the parameter range, number of objects in each bin, median redshift, 24{\um}, 100{\um}, and 160{\um} stacked flux densities and their measurement uncertainties, rest-frame 7.7{\um} luminosity and its measurement uncertainty, {\lir} and its uncertainty, and ratio of stacked $L_{7.7}$ to stacked {\lir} and its error.}
	\tablenotetext{*}{Stacks are undetected in 24{\um}, 100{\um}, and 160{\um}, and hence, $L_{7.7}$ and {\lir} are 3$\sigma$ upper limits. {\ratio} is meaningless in this case.}
\tablenotetext{a}{{\lir} is derived from $f_{100}$ and $f_{160}$ fit to CE01 templates. Its error is the standard deviation of 10,000 {\lir} realizations that are calculated by fitting IR templates to the perturbed $f_{100}$ and $f_{160}$. If both $f_{100}$ and $f_{160}$ are undetected we used upper limits, but if only one is undetected we used the flux and its error to find the best-fit model and the associated {\lir}.}
	\tablenotetext{b}{N2 stacks are limited to objects at $2.0\leq z\leq 2.6$, because otherwise the highest metallicity bin is dominated by low-$z$ galaxies with $\hat{z}\sim 1.6$. The rest of the stacks (O3N2, {\o32}, $M_*$) include all objects at $1.37\leq z\leq 2.60$. The median redshifts in these bins are consistent and are above 2. We should note that {\o32} sample has only 8 galaxies with $z<2.0$.}

\label{tab:pah_ir}
\end{deluxetable*}
\capstarttrue  

\end{document}